%% file: memoir_M2_arXiv_sub.tex
\documentclass[12pt,a4paper]{article}
\usepackage[utf8]{inputenc}
\usepackage{natbib}
\usepackage[english,francais]{babel}
\usepackage{amsmath}
\usepackage{amsfonts}
\usepackage{amssymb}
\usepackage{makeidx}
\usepackage{graphicx}
\usepackage{url}
\usepackage{float}

\usepackage{hyperref}
\hypersetup{
    colorlinks=true,
    linkcolor=blue,
    filecolor=blue,      
    urlcolor=blue,
    citecolor=blue,
} 
\usepackage{multirow}
\usepackage{array}
\newcolumntype{L}[1]{>{\raggedright\let\newline\\\arraybackslash\hspace{0pt}}m{#1}}
\newcolumntype{C}[1]{>{\centering\let\newline\\\arraybackslash\hspace{0pt}}m{#1}}
\newcolumntype{R}[1]{>{\raggedleft\let\newline\\\arraybackslash\hspace{0pt}}m{#1}}
\usepackage{arydshln}
\usepackage[font=footnotesize, labelfont=bf]{caption}
\usepackage{lipsum}
\usepackage[font=scriptsize, labelfont=bf]{subcaption}
\usepackage{fullpage}

\usepackage{enumitem}

\usepackage[left=2.5cm,right=2.5cm,top=1.75cm,bottom=1.5cm]{geometry}
\author{Adrian BACH}
\title{Internship memoir}

\begin{document}

\input{memoir_titlepage.tex}

\section*{Abstract}

To date, the mechanisms underlying the diversity of the emergent patterns of collective motion in locust hopper bands remain to be unveiled.
This study investigates the role of speed heterogeneity in the emergence of the most common patterns (frontal and columnar), following the Self-Organization framework.
To address whether marching activity intermittency and density-dependant hopping individual behaviours could underlie the formation of such patterns, a three-zone Self-Propelled Particles model variant was formulated. In this model, individuals alternated between 
marching and resting periods, and were more likely to hop when crowded.
Simulations of half a million locust nymphs marching during eight hours, in fifty replicates of more than five hundred parameter combinations, were ran on General Purpose Graphical Processing Units, using the CUDA C++ language for parallel computing.
The model successfully predicted the emergence of both patterns of interest, with the presence of a density-dependent hopping probability being a necessary condition. Short to absent pause periods mostly resulted in columnar shapes, similar to the ones observed in the brown locust, \textit{Locustana pardalina}, and long pause periods rather resulted in frontal shapes, such as exhibited by the Australian plague locust, \textit{Chortoicetes terminifera}.
Furthermore, the density profiles of simulated frontal formations displayed the same frontal peak followed by an exponential decay as empirical profiles of Australian plague locust hopper bands.
Both simulated and experimental paint marking experiments showed that locusts initially located at different positions in the band were find together at its front after a few hours marching; an expected global behavior in hopper bands undergoing activity intermittency.
These results represent an important first step towards a cross-species comparison of locust mass migration patterns.

\section*{Résumé}

Les mécanismes à l'origine de la diversité des structures de mouvement collectif chez les bandes larvaires de criquets n'ont, à ce jour, pas encore été révélés.
Dans cette étude, nous explorons le rôle de l'hétérogénéité de la vitesse dans l'émergence des formes frontales et en colonnes, dans le cadre de la théorie d'Auto-Organisation.
L'intermittence de l'activité et le saut densité-dépendant pourraient-ils être à l'origine de la formation de ces structures? Pour y répondre, une variante du modèle des Self-Propelled Particles (particules automotrices) à trois zones a été formulée. Dans ce modèle, les larves alternaient entre des phases de marche et de pause, et étaient plus susceptibles de sauter lorqu'elles étaient très densément entourées.
Un demi million de larves marchant pendant huit heures a été simulé pour cinquante réplicats de plus de cinq-cent jeux de paramètres, grâce à une implementation en CUDA C++, un langage créé pour le calcul en parallèle sur GPGPU (General Purpose Graphical Processing Unit). 
Le modèle a pu prédire l'émergence des deux formes d'intérêt et la prise en compte du saut densité-dépendant était une condition indispensable.
Une période de pause courte, voire absente, a favorisé la formation de colonnes semblable à celles du criquet brun (\textit{L. pardalina}), tandis qu'une longue période produisait des formes frontales du type du criquet australien (\textit{C. terminifera}).
D'autre part, les profils de densités issus des simulations ayant donné des fronts étaient structurés en un pic suivi d'une décroissance exponentielle, similaire à celle des profils empiriques mesurés sur des bandes larvaires de criquet australien.
Les expériences de marquages sur le terrain, ainsi que leur simulation, ont montré que des criquets initialement positionnés à différent endroits de la bande étaient tous retrouvés en même temps au front après quelques heures de marche; un comportement collectif attendu en présence d'une intermittence de la marche.
Ces résultats sont une première étape importante vers une étude comparative approfondie des formations migratoires de différentes espèces de criquets.

\newpage
\tableofcontents
\newpage


\section{Introduction}
\begin{figure}
\centering
\includegraphics[scale=0.7]{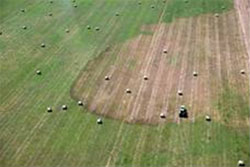}
\includegraphics[scale=0.19]{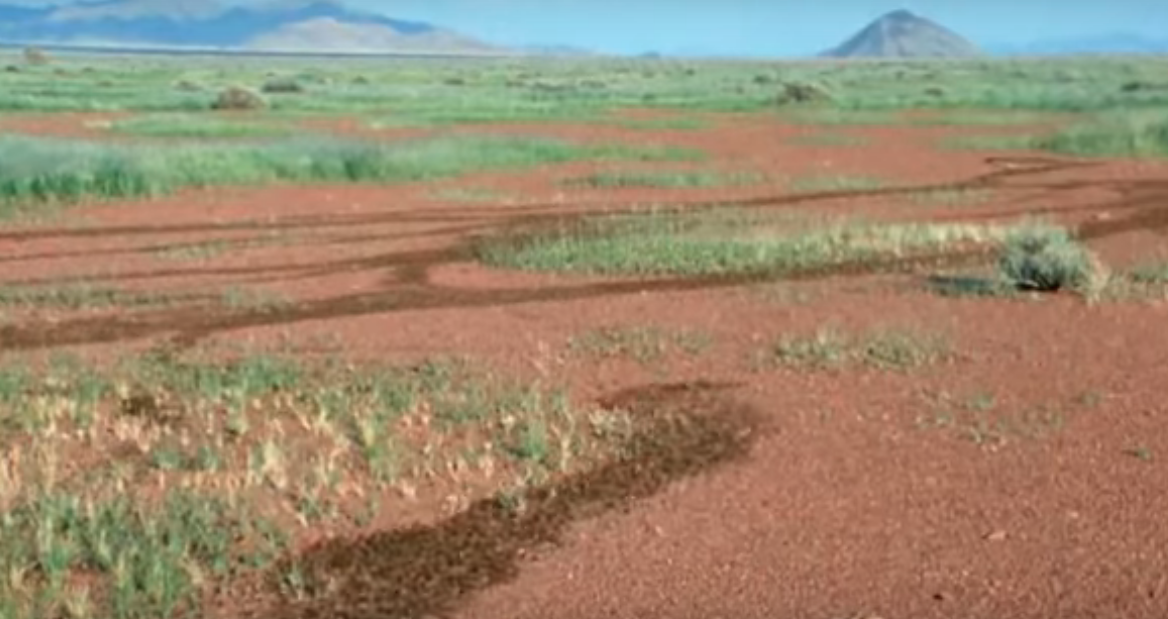}
\caption{Left panel shows an Australian plague locust hopper band in frontal formation progressing from the upper right corner to the lower left one (source : Hunter), right panel shows a a brown locust hopper band in columnar formation progressing from the left to the right side of the image (source : BBC video).}
\label{bands}
\end{figure}

In the context of pattern formation in biological systems, \cite{camazine2003self} defined Self-Organization (SO) as \textit{"a process in which pattern at the global level of a system emerges solely from numerous interactions among the lower-level components of the system. Moreover the rules specifying interactions among the system's components are executed using only local information, without reference to the global pattern."}
This feature is a spectacular source of variability as, in self-organised systems, the smallest change in the interaction rules can lead to dramatic modification of emerging patterns.\\
In biology, SO appears to be involved in systems of all scales, from cells combining into structured tissues to pigmenting patterns in animal coats, and also a wide range of collective behaviors.
Living in groups provides animals with interesting evolutionary advantages, such as protection, dilution of predation, more efficient foraging or even improved decision making.
In mobile social animals, selection has shaped individual-level mechanisms allowing groups to maintain these advantages when moving. 
Such a self-organized collective motion usually results in groups of animals cohesively moving in the same direction without following a leader nor any external constraints, such as ones originating from topology or wheather \citep{camazine2003self}.\\
To understand the emergence of a biological pattern through SO, individual-based models often constitute a useful key hypothesis testing tool, as they can reveal the global-level consequences of a given set of lower-level rules using simulations.
In self-organized collective motion, a group is usually modelled as a set of moving individuals, similar in terms of shape and capacities, that share common local interaction rules, implemented and parameterized on the basis of behavioural experiments.
If the model successfully predicts the global-level pattern observed in nature, it is a strong argument in favor of the rules it was built on.
Among these models, the Self Propelled Particles (SPP) model stood out due to its few, simple and non-specific rules, namely adjusting a particle's direction to the one of close neighbours with a certain error \citep{vicsek1995novel}.
Initially designed for the investigation of the SO-typical phase transition from order to disorder in moving particles according to noise and density \citep{toner1998flocks, gregoire2003moving, nagy2007new}, its flexibility to extra hypotheses was paramount in understanding the local rules behind many animal collective motion, such as fish schools \citep{gautrais2012deciphering}, or starling flocks \citep{ballerini2008interaction} to cite the most common examples.
\cite{buhl2006disorder} recently showed that locust migrating groups also exhibited this non-linear phase transition from disordered movement to collective motion when increasing density, which had been well predicted by SPP simulations.\\
Locust migrations have a bad reputation because the crop-devastating swarms they form when dense groups of adults fly together.
Yet, before they fledge, locusts march in hopper bands composed of up to millions of wingless nymphs, stretching over hundreds of meters, sometimes kilometers.  \citep{uvarov77}.
\cite{buhl2012using} showed that the local interactions responsible for the maintainance of this collective motion were a balance of repulsion in a close range, alignment in an intermediate range and attraction to conspecifics in a larger range.\\
Locust nymphs display a spectacular range of migratory band patterns, from \textit{frontal} to \textit{columnar} formations (figure \ref{bands}).
The frontal formation was described as a comet-like shape with a very dense linear front, and a density-decaying, more scattered tail that can be observed in various species, including the Australian plague locust \textit{Chortoceites terminifera} (APL) \citep{buhl2011group}.
The columnar formation is a network of dense and narrow interwined streams, seen in the South African brown locust \textit{Locustana pardalina} (\cite{uvarov77}, see also \url{https://www.youtube.com/watch?v=1YNy2R3hg2Q}).
This variety of shapes could not be predicted by a SPP based on the previous hypotheses, and the mechanisms at the origin of their emergence remain to be studied \citep{sumpter2010collective}.\\

In fish schools, the emergence of either oblong or frontal shape has already been shown to be linked to speed heterogeneity \citep{hemelrijk2008self}.
Interestingly, several field studies previously reported that hopper bands were considerably slower than the average speed of the nymphs comprising it, which could be a consequence of a speed heterogeneity among the individuals. 
This feature could be induced by the alternance between walking, resting and hopping exhibited by locust nymphs \citep{uvarov77}.
Previous studies on feeding patterns and short endogenous rythms by \cite{simpson1981oscillation} and \cite{SIMPSON1986480} showed that \textit{L. migratoria} presented an hourly interval between feeding activities and pauses of at least 15 minutes linked to feeding-related behaviour.
This information led to the following hypothesis on the structure of locust nymphs migrating bands.\\
As locusts move faster than the band, actively marching individuals would quickly reach the front, forming a dense leading edge, and then rest for a variable period of time.
During this pause, the edge would continue its way over them, leaving them behind.
Therefore, the longer they stay inactive, the further away at the back of the band they would find themselves.
After pausing, these individuals would actively march to the leading edge again, and so on, forming a rolling structure.
This variability in the time spent inactive would result in locusts resting at many different positions at the back, but always gathering to the front when active.
As a consequence of this, the edge would be denser than the back, where locusts would be distributed in a more scattered configuration, which matches the description of a frontal formation.\\
Nymphs of all locust species tend to alternate between walking and hopping when on the move \citep{uvarov77}, and APL hop visibly more frequently than other well studied species. Buhl (personal communication) hypothesised that hopping probability could be higher when the hopper is very densely surrounded, possibly to avoid rear contact with pairs that could result in cannibalistic attacks \citep{bazazi2008collective}. Hopping being another speed heterogeneity-inducing behaviour, it could also have a role to play.

Could intermittency and density-dependent hopping behaviours explain the diversity of hopper bands formations? If so, what is their role, along with the social interactions interplay, in the emergence of the different migratory band shapes?
As it is difficult to anticipate precisely what effects a combination of intermittency in marching activity and hopping might have on pattern formation without simulations, Buhl implemented a SPP model variant, in which these new hypotheses were added to the social interaction rules of repulsion, alignment and attraction.\\
I used Buhl's new model to simulate half a million locust marching during several hours for a wide range of parameter sets, exploiting GPGPU (General Purpose Graphics Processing Unit) parallel computing power of the University of Adelaide super-computer Phoenix.
I then arranged and displayed the models outputs in different configurations to show the effect of these local rules, and qualitatively assess the role of intermittency and crowded hopping on the emergence of frontal or columnar patterns. I also computed a set of quantitative measures of band shape from simulations as a first step towards a more in depth data analysis.\\
To validate these results I confronted the model predictions to actual locust nymph bands behaviour, using two experiments from Buhl's team former field studies on APL hopper bands.
The first experiment was a count of the number of locusts within a fixed area set in the way of the bands, which permitted to plot locust density according to time (density profiles). Analyses from \cite{buhl2011group} showed that these profiles displayed a sharp and intense peak at the front, followed by an exponential decay from the peak towards the back.
The second was a preliminary experiment to validate the hypothesis of the rolling structure.
If it was the case, individuals initially at the front would find themselves at the back after resting, meanwhile an active individual initially at the back would reach the front.
Thus, after a few hours marching, one should be able to find locusts of any initial position all together at the front.
To test for this, the team marked a large number of nymphs with paint according to their positions in the band, and counted them after allowing the band to march for a few hours.
I reproduced these experiments \textit{in silico} on model outputs and assessed if the simulations that resulted in frontal formations (such as APL) exhibited the same structure, assessing that the model had succesfully predicted the band behaviour.

\newpage
\section{Material and method}
\label{matmet}

\subsection{Model}
\label{model}

There are four locusts collective motion models in the literature that I reviewed : the escape and pursuit model based on cannibalistic interactions by \cite{romanczuk2009collective}, the metric-SPP model by \cite{buhl2012using}, the pause-and-go model by \cite{ariel2014individual}, based on the questionable assumption that locusts change direction only when stopped, and the Alignment and Intermittent Motion model (AIM) in \cite{jones2016hopper} as a mix of the latter two.
For now, the metric-SSP model has the strongest experimental and empirical basis, and has been shown in \cite{buhl2012using} to be better fitted to field measures than the escape and pursuit model.
We adapted this model to the hypotheses presented further up and optimized it for parallel computing. 

\subsubsection{Three-zone SPP}

In our model, an individual $i$ was characterized by $\vec{r_i^t}$ , its ($x;y$) position in a 2D space at time $t$, and $\vec{\theta_i^{t}}$ its ($\theta_x;\theta_y$) normalized direction at time $t$. Note that direction has been implemented as a vector rather than an angle for efficiency purpose. Direction was normalized so it did not influence the sum of forces presented further.
At each time step $\Delta t$, each individual moved at a variable speed $v$ (can be 0, $v_0$ or $v_0 + H$, see next sextions) and its next position was computed as follows:

\begin{equation}
\label{eq1}
\begin{aligned}
\overrightarrow{r_i^{t + \Delta t}} &= \overrightarrow{r_i^t} + v\; \Delta t\ \overrightarrow{\theta_i^{t}}'
\end{aligned}
\end{equation}

When moving, an individual chose its direction with a certain error modelled by a random angular variable $\hat{\xi_i}$ drawn in a uniform law between $[-\pi;\pi]$, weighted with a positive parameter $\eta$.

\begin{equation}
\label{eq2}
\begin{aligned}
\overrightarrow{\theta_i^{t}}' &= (1 - \eta)\overrightarrow{\theta_i^{t}} + \eta\left(\begin{array}{c}\cos{\hat{\xi_i}}\\ \sin{\hat{\xi_i}}\end{array}\right)
\end{aligned}
\end{equation}

Each particle moved according to a correlated random walk. In this type of walk, individuals next direction is influenced by inertia, so $\vec{\theta_i}$ was proportional to its former value with a positive coefficient $\sigma$. Beside inertia, next direction depended on the social interaction component $\vec{\omega_S}$, a unit vector of coordinates ($\omega_{S,x}\:$;$\:\omega_{S,y}$). Thus, at each time step, the next direction was computed as follows :

\begin{equation}
\label{eq3}
\begin{aligned}
\overrightarrow{\theta_i^{t + \Delta t}} &= \sigma\overrightarrow{\theta_i^{t}}' + (1 - \sigma) \overrightarrow{\omega_S}
\end{aligned}
\end{equation}

The social interaction component $\vec{\omega_S}$ resulted from normalizing $\vec{\Omega_S}$, the vectorial sum of social, pairwise interaction forces $\vec{f_{ij}}$ exerced by a neighbour $j$ on focal individual $i$ within its perception range $R_p$. This range was divided in three concentric sections.
\begin{itemize}
\item Repulsion range, of $R_r$ radius, was a spatial region in which the focal individual seeked to avoid contact with others in this range by moving in their opposite direction.
\item Alignment range, between the $R_r$ and the $R_a$ radii circles, in which focal individual aligned its direction with the average of others' direction within this range.
\item Attraction range, between the $R_a$ and the $R_p$ radii circles, in which focal indidual aimed to go towards the others within this range.
\end{itemize}
Following these rules, interaction forces were computed this way :

\begin{equation}
\label{eq4}
\begin{aligned}
\overrightarrow{\Omega_S} &= \sum_{j} \overrightarrow{f_{ij}} \\
\vec{f_{ij}} &= \left\{
                \begin{array}{lll}
                  -\alpha.\overrightarrow{e_{ij}} &if &r_{ij} \leq R_r\\
                  \beta.\overrightarrow{\theta_j^t} &if &R_r < r_{ij} \leq R_a\\
                  \gamma.\overrightarrow{e_{ij}} &if &R_a < r_{ij} \leq R_p\\
                \end{array}
              \right.\\
\end{aligned}
\end{equation}

With $\alpha$, $\beta$, $\gamma$ respectively repulsion, alignment, and attraction weight parameters. $\vec{e_{ij}}$ was the unit vector from individual $i$ towards $j$. 
Note that this way to compute interaction forces was slightly different than the one in \cite{buhl2012using} which involved a alignment force within $R_a$ and a Lennard-Jones force type to balance repulsion and attraction according to the distance to focal individual. But it coupled attraction and repulsion with a single weight parameter, and attributing a weight to each different interaction type allowed more freedom in the investigation of the role of attraction and repulsion independently.\\

Inter-individual interactions have been implemented in various ways in previous studies, including using fixed radii determining what type of interaction occurs according to distance ("metric" method; \citep{gregoire2003moving, hemelrijk2008self}), or the neighbours' nearest distance rank ("topological" ranges, \citep{ballerini2008interaction, gautrais2012deciphering}) which is more relevant when perception distances are not limiting. We chose a metric implementation of interaction ranges which fits well with empirical data \citep{buhl2012using} and the fact that locust have a very limited range of perception.\\

Physical collisions between individuals were not considered, as locusts nymphs are able to march over each other with ease at high densities, sometimes leading to several layers.

\subsubsection{Hopping}
\label{hop}

In a 2D context, a jump can be seen as an instantaneous and temporary increase in marching speed. In \cite{buhl2012using} simulations, at each time step a moving individual had a probability $p_h$ to spontaneously hop. When it did, it moved with an additional speed $H$ for this time step only.
One of the novelties was a higher probability to hop if the individual was crowded. A hopper was considered crowded when there was at least one conspecific in its $R_r$ range, in this case, the individual had another, greater probability to hop $p_{hc}$ for this time step only.

\subsubsection{Marching activity intermittency}

In our model, each locust had a fixed marching activity period of $A$ time steps, after which they stopped for a fixed period of $I$ steps. Beyond this minimal inactivity period, each locust had a constant probability $p_{res}$ to resume marching per time step, which represented the variability in the time each locust spent inactive.
It was implemented as an individual time tracker starting at $A$ that was subtracted by one at each time step. An individual marched when its value on the time tracker was from $A$ to 0, it stopped marching ($v$ is set to zero) from 0 to $-I$ and after this value, the tracker could be reset to $A$ with a $p_{res}$ probability at each time step. 

\subsubsection{Occlusion}

In the situation of high densities around an individual, as the attraction range covered a larger surface than repulsion and alignment ranges, attraction force could have too much weight in the sum of forces because the potential number of individuals in it was higher. It does not reflect collective motion on the field, as it was unlikely that such distant pairs influenced an individual's estimation of direction because of the visual occulsion by others within its first two ranges \citep{hemelrijk2008self}. This artefact was removed by setting a threshold number of individuals within $R_a$, above which the attraction did not contribute to $\vec{\Omega_S}$ computing anymore.

\subsubsection{Implementation}

The world was a grid of 512 x 512 squared cells equal to twice the attraction range.
This system was used to optimise data sorting during parrallel computation because, for any particule within a cell, all the potential interacting neighbours could be found in the directly adjacent cells.
The total length was approximately 307,2m, which was coherent with the daily distance an actual band would travel.
Thus, it was unlikely that the band reached a boundary, exept at the start of simulation (based on previous preliminary exploration).
The boundaries were chosen to be reflective as the 2D world represented a flat clear ground, and that an individual crossing a boundary was not expected to appear at the opposite side (which is the case of periodic boundaries).\\
At the initialization of the simulation, the positions of individuals were randomly distributed in a round shaped defined set against the left boundary and half-way between the upper and lower boundaries. This area was set so that locust density was 750/m2 (in this case, a radius of $r_i=$14.92m, with the centre of the circle at coordinates $x_i=$14.92m and $y_i=$153.6m), to ressemble hopper bands basking at dawn \citep{uvarov77}.
Since marching activity occurs after a certain amount of time allowed to other activities (perching, feeding, \ldots), each individual's initial value in the time tracker was drawn in a uniform law between $A$ and $-I$.\\
Each individual's initial direction was set towards world's right side with a noise drawn in a uniform law between $[-\pi;\pi]$, so that the group was initiated with a high degree of collective alignment, as locusts would be after a short period of time during the onset of marching.\\
The simulation time was 480min (time step $=$ 1s) to reproduce a typical maximal daily marching activity period among locust nymphs \citep{uvarov77}.
Every 5 minutes, the state of the 2D map was recorded in the output text file.\\
Among all individuals, 3 groups of 1000 locusts had their positions specifically set in a small area, either at the front, the middle, or the back of group, for the purpose of comparing their movement within the band with empirical data on paint-marked individuals. The middle group was randomly distributed within a circle of radius$=$0.05 x $r_i$ with the same centre as the rest of group (at coordinates $x_i$ and $y_i$), while the front and back group where randomly distributed within an area at a distance between 0.95 x $r_i$ and $r_i$ from the group initialisation centre (at $x_i,y_i$) and with an angle between -1/15 x $\Pi$ and 1/15 x $\Pi$ for the front group and 14/15 x $P_i$ and 16/15 x $P_i$ for the back group.

\subsubsection{Fixed parameters}
\label{param}

The time step $\Delta t$ has been set to one second, as it was a good trade-off between simulation time and computing speed on GPGPUs. A total of $N = 524288$ $(2^{19})$ particles were simulated.
Locusts march at a slow speed of 2.5 mm.s$^{-1}$ speed between bouts of jumping (Buhl, unpublished data).
Previous model exploration for \cite{buhl2006disorder} and \cite{buhl2012using} simulations have led to set the correlation factor between former and next direction $\sigma$ to 0.6.
\cite{buhl2011group} have shown that in Australian plague locust $R_a$ is around 13.5 cm, and \cite{buhl2012using} that $R_r$ is around 3 cm. Results from \cite{simpson1999behavioural} and \cite{gray2009behavioural} suggested that locusts a have visual perception range $R_p$ limited to 30cm, beyond which they are not attracted to conspecifics anymore.
The noise intensity $\eta$ was set to 0.05 as previously used in small scale 2D simulations \citep{buhl2012using}.
The hopping probablility was set to 0.01, and the hopping speed bonus $H$ to 10 cm.s$^{-1}$ (Buhl, unpublished data).
\cite{simpson1981oscillation, SIMPSON1986480} showed that there tends to be an interval of 60min between feeding bouts in locusts, with feeding pauses generally lasting 15min and a constant probability to resume marching per time step afterwards, until the next feeding pause 45min later. Such 60 min rythmic cycles were also apparent in laboratory experiements involving marching activity in the APL (Buhl and Simpson, personnal observations).
Thus, the activity period $A$ was set to 2700 time steps, which are 45 minutes.
The number of conspecifics in the alignment range to create an occulusion $Oc$ has been set to 25 individuals, an intermediate value of the number of locusts that could fit in the alignment range, beyond which it becomes difficult for further locusts to remain visible.
See table \ref{fixed} for a summary.

\begin{table}[h]
\centering
\footnotesize
\caption{Fixed parameters values. \textsf{(u)} stands for unitless.}
    \vspace{0.15cm}
    \sffamily
\label{fixed}
\begin{tabular}{C{5cm} C{2cm} C{2cm} C{5cm}}
	Parameter & Notation & Value & Source \\ \hline \\
	Average speed & $v_0$ & 0.0025 m.s$^{-1}$ & (Buhl, unpublished data) \\ \cdashline{1-4} \\
	Weight of previous direction & $\sigma$ & 0.6 (u) & \cite{buhl2006disorder, buhl2012using} \\ \cdashline{1-4} \\
	Weight of random component & $\eta$ & 0.05 (u) & \cite{buhl2006disorder, buhl2012using} \\ \cdashline{1-4} \\
	Repulsion range & $R_r$ & 0.035 m & \cite{buhl2012using} \\ \cdashline{1-4} \\
	Alignment range & $R_a$ & 0.135 m & \cite{buhl2011group} \\ \cdashline{1-4} \\
	Attraction range & $R_p$ & 0.300 m & \cite{simpson1999behavioural, gray2009behavioural}\\ \cdashline{1-4} \\
	Hopping probability & $p_h$ & 0.01 (u) & (Buhl, unpublished data) \\ \cdashline{1-4} \\
	Hopping speed bonus & $H$ & 0.1 m.s$^{-1}$ & (Buhl, unpublished data) \\ \cdashline{1-4} \\
	Activity period & $A$ & 2700 s & \cite{simpson1981oscillation, SIMPSON1986480} \\ \cdashline{1-4} \\
	Occlusion threshold & $Oc$ & 25 individuals & (Buhl, personal observation) \\ \hline
\end{tabular}
\end{table}

\subsubsection{Systematic exploration}

To select the parameter ranges to be explored, we ran a preliminary exploration of the model with only one replicate per parameter set to view a wide range of the possible patterns. Then, we qualitatively filtered for the parameter values that produced the same patterns as others or produced shapes too irrelevant to this study. We finally opted for the parameter ranges presented in table \ref{tuning}.
We treated the minimal inactivity period and the probability to resume walking as couples because they both influenced the same feature, being the mean period of pauses.
There were 576 different parameter sets for the final exploration.\\
I ran fifty replicates per parameter set as it was a good trade-off between statistical relevance and computing labor.

\begin{table}[h]
\centering
\footnotesize
\caption{Free parameters ranges for exploration.}
    \vspace{0.15cm}
    \sffamily
\label{tuning}
\begin{tabular}{C{3cm}|C{3cm} C{3cm} C{4cm}}
	Type & Parameter & Notation & Values \\ \hline \\
    \multirow{3}{*}{Social component} & Repulsion weight & $\alpha$ & 0.1; 1; 1.5 \\ \cdashline{2-4} \\
    									 & Alignment weight & $\beta$ & 0.1; 1.0; 3.0 \\ \cdashline{2-4} \\
    									 & Attraction weight & $\gamma$ & 0; 0.001; 0.01; 0.05 \\ \hline \\
    									 
	\multirow{3}{*}{Speed heterogeneity} & Crowded jumping probability & $p_{hc}$ & 0; 0.1; 0.2; 0.3 \\ \cdashline{2-4} \\
                  				 		& Minimal inactivity period and probability to resume marching couples & $I/p{_res}$ & 0/1; 90/0.5; 900/0.001; 2700/0.0005 \\ \hline 
\end{tabular}
\end{table}

\subsection{Simulations}
\label{simu}

All computational operations required for a single individual were rather simple but
if they were computed sequentially, the simulation time would increase non-linearly with the number of particles simulated. When studying groups of several thousands individuals, the limiting factor used to be computer throughput, meaning the number of computations being able to be realized simultaneously. Parralel computing uses a large number of series small Central Processing Unit, that permits performing many simple computing tasks simulatneously. Recent advances in General Purpose Graphical Processing Units (GPGPUs) have permitted to enhance and optimize both computing speed and throughput, allowing the simulation of systems containing up to millions of individuals in a considerably shorter time.\\

For this natural-scale study, one of the main challenges was the order of magnitude of the number of particles to be simulated.
To simulate half a million of them, the model was implemented in the CUDA C++ language to exploit GPGPU parallel computing power.
As I needed several replicates for several parameter sets in a relatively short time, I used high performance computing to run a large number of simulations simultaneously.
The University of Adelaide made it possible by allowing me access to their super-computer Phoenix. 
Phoenix was recently ranked among World's top 500 most powerful super-computers list, with 7468 cores, 336 GPU accelerators and 45 TB memory for a 450 Tflops computing capacity.
All the simulations were ran on Phoenix' GPU partition.

\subsection{Empirical data}
\label{data}

All empirical data concerning hopper band density profiles were previously published in \cite{buhl2011group}.
Density profiles were quantified by analysing video recording of locusts moving under a tripod mounted camera set ahead of the band's front (for further details, see reference). 
Tagging experiments were preliminary trials performed on hopper bands found near Hillston, NSW, in November 2010.
When possible, large groups of individuals (several thousands) were painted using two different colours.
One colour was chosen to mark a group of locusts captured at the front of the band, while the other was used on locusts captured 100 meters away rearwards. Captured locusts were painted using a water based and non toxic fluorescent paint in pink or yellow, applied as a fine mist using a handheld sprayer.
They were released in the same location as their capture and the band was left to move for a further 3h20 before observation.
Each observer was positioned at a fixed location where they measured a square meter area on which they counted the presence of marked locusts of each colour totalled every 5 minutes for 2 hours.
Observers were all located towards the middle of the front of the band, spaced 5m apart. 
To reproduce that protocol in the simulations, a tag number was assigned to a large subset of particles according to their position at initialization (front, middle and back).

\subsection{Measures and statistics}
\label{stats}

\subsubsection{Density profiles, band's size and trajectometry measures}

To compare simulation predictions to field data,
I measured the density of locusts crossing a line (a thin vertical band of 1x512 cells) that was set ahead of the motion of the group, in a similar way to how empirical data was gathered.
I drew this line at a time step when I assumed the collective movement to be settled. I chose fifty minutes as it was a time where most columns and fronts formations appeared well estabished. To set the line as tangentially as possible to the front, I changed the coordinate system to a ($\vec{e_x'}$;$\vec{e_y'}$) base in which $\vec{e_x'}$ unit vector was colinear to the mean direction of the band at each time step. Then, I fixed the measure line one cell away from the most frontal cell in $x'$ direction.\\
I also analysed the simulations' output to quantify several measures of the band cohesion and shape.
I extracted from each simulations at each time step the average direction vector of the band (the magnitude of this vector, often called "polarisation" or "alignment" denotes the degree of collective alignment in the group), the number of occupied cells, the center of mass coordinates (before the change of coordinate system) and density of nymphs in the measure band (in total and for each tag separately).
At the time steps we decided to print the simulations out (240min and 480min), I computed band's length, width, most frontal and most rearward locust.
I averaged all these measures over the fifty replicates and computed standard deviation and 95\% confidence interval.
This is a preliminary analysis that will be followed with further in depth analysis of the shape and trajectories of bands.
All these operations were ran on Phoenix CPU partition using the Perl language.

\subsubsection{Parameter exploration and sensitivity analysis}

I could not find an existing multivariate sensitivity analysis package available for non-deterministic models such as ours. I considered a Co-Inertia Principal Component Analysis between parameters and shape measures tables, but it did not seem to be applicable to our dataset. 
As a preliminary statistical approach, we ran a Spearman's non-parametric correlation test between each parameters values and shape measures independently using MatLab.\\
In addition to those preliminary statistics, I represented the results graphically in order to qualitatively explore the effect of the model parameters on locust band patterns.
After checking that patterns did not strongly vary between replicates within a given parameter set, I printed out  one replicate of each parameter set after four and eight hours of marching (using Perl GD package and ImageMagick), which is the maximal daily marching activity.
To illustrate how the model behaved without our new intermittency and hopping rules, I printed out one simulation per parameter sets with $p_{hc} = 0$ and $I=0$ / $p_{res} = 1$ couple after four hours of marching and arranged them in a $\alpha$, $\beta$, $gamma$ parameter landscape. In comparison, I printed out, one simulation per parameter sets with non-zero values for $p_{hc}$ and $I/p_{res}$ couple after four hours of marching and arranged them in same parameter landscape.\\
To assess the role of intermittency in the emergence of either frontal or columnar patterns, I printed out one simulation per parameter set for fixed values of $gamma$ and $p_{hc}$ and arranged them in an $\alpha$, $\beta$, $I/p_{res}$ couple parameter landscape. Similarly, to investigate the role of $p_{hc}$, I printed out one simulation per parameter set for fixed values of $gamma$ and $I/p_{res}$ and arranged them in an $\alpha$, $\beta$, $p_{hc}$ parameter landscape.
As a quantitative back up for visual analysis, I computed for each parameter sets in question the average over the fifty replicates of bands' surface (in km$^2$), elongation (length to width ratio) and mean alignment after four hours marching and plotted them in the same parameter lanscapes using R.

\subsubsection{Comparison to field data}

If the model fitted field data on APL, we would expect a sharp peak with a steep increasing at the most frontal position of the band, followed by a exponential decay in direction of the rear \citep{buhl2011group}.
To test for this, I qualitatively selected the parameter sets resulting in APL-like, frontal-shaped bands, checking that they kept the same shape even after eigth hours of marching (24 candidates). Afterwards, I plotted the associated density profiles, and qualitatively assessed the effect of $p_{hc}$ and $I/p_{res}$ couple on their structure. I applied linear regressions on the logarithm of densities after the peak occurence (using R) to ensure the decay was exponential, and computed the mean R$^2$, the average time of occurence of the peak after the first locust counted, and the average length of the bands to compare them to available field data.\\
In frontal formations, we suspect that marching activity intermittency results in active individuals at the back of the band quickly reaching the front, while inactive individuals at the front might find themselves further back during long pauses.
If so, we would expect individivuals initially from different locations in the band to be observed together at the front after a sufficient amount of time spent marching.
As a preliminary approach, I compared the tag-specific density profiles from a simulated frontal formation to one measured on the field.

\newpage
\section{Results}

\subsection{The need for new hypotheses}
\label{res1lab}

\begin{figure}
    \begin{minipage}[c]{.46\linewidth}
        \centering
        \includegraphics[scale=0.5]{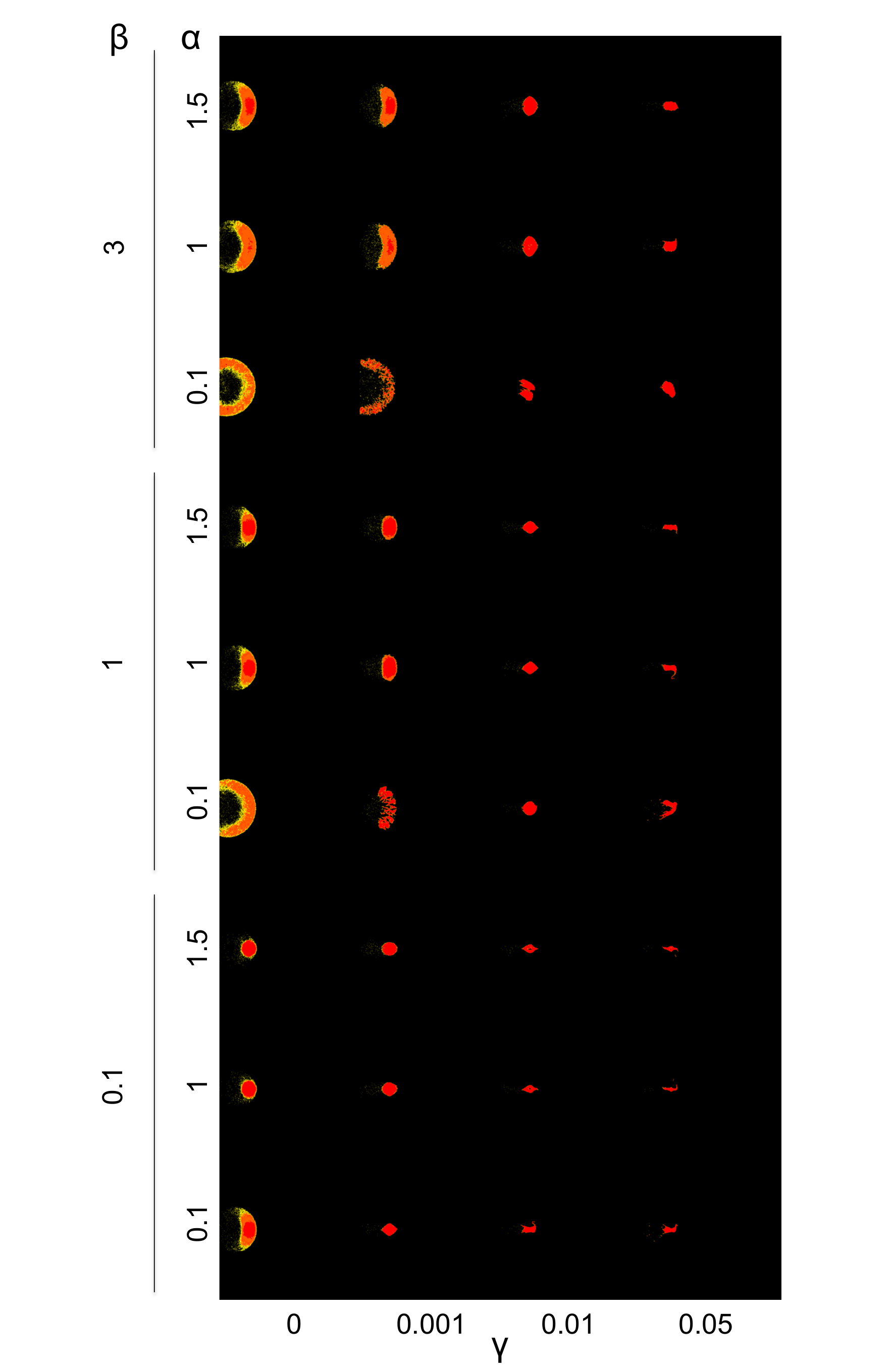}
    \end{minipage}
    \hfill%
    \begin{minipage}[c]{.54\linewidth}
        \centering
        \includegraphics[scale=0.5]{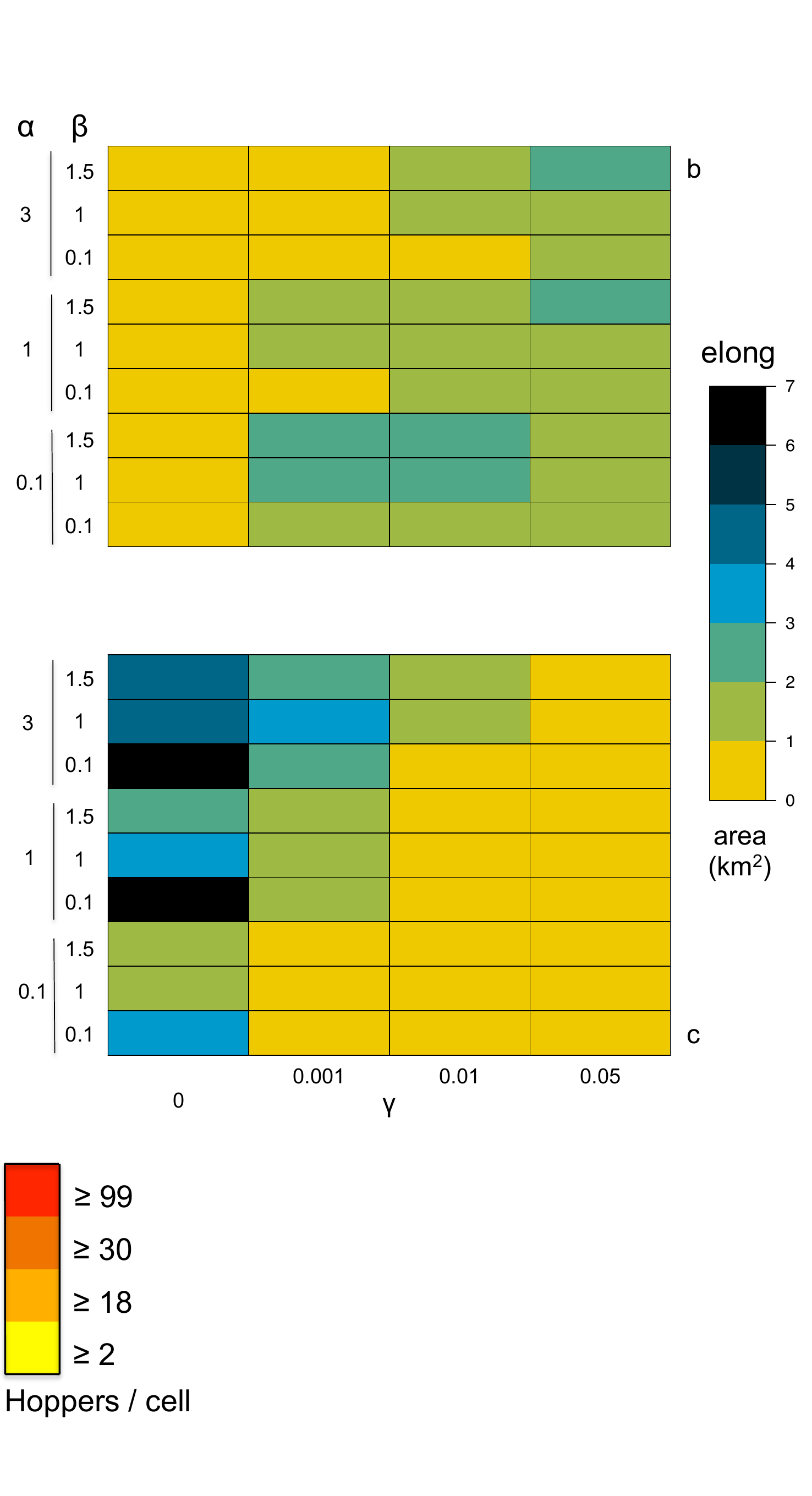}
    \end{minipage}
\caption{Parameter landscapes for \textsf{(a)} the simulated bands after four hours marching (colours describe hopper density), \textsf{(b)} bands' elongation and \textsf{(c)} area covered for $p_{hc} = 0$ and $I=0$ / $p_{res} = 1$ couple.
Left vertical axes are repulsion weight $\alpha$ in the individual interaction force computing, the right vertical ones are alignment weight $\beta$ and the horizontal are the weight of attraction $\gamma$.
The bands elongation and area are presented on the same scale.
Note that elongation can be misleading in the case of cylindric shapes.
The white arrow indicates the direction of the band at initialization.
The shapes were not qualitatively different after eight hours marching.}
\label{res1}
\end{figure}

This section illustrates the patterns that emerged when neither intermittency nor crowded hopping rules were taken into account in the model ($p_{hc} = 0$ and $I=0$ / $p_{res} = 1$ couple, 36 parameter sets, see figure \ref{res1}). It corresponded to locusts marching without stopping and all having the same probability to hop.\\
The simulated hopper bands for these values displayed a higher variability in width (55.50 m $\pm$ 36.54 SD; n$=$36) than length (54.87 m $\pm$ 13.84 SD; n$=$36). It was consistent with previous studies showing that "classical" SPP have a more marked diffusion in the transversal axis  compared to the longitudinal axis (in relation to group movement) \citep{toner1998flocks}.
Most of the bands were rather round-shaped (mean elongation : 1.32 $\pm$ 0.61 SD; n$=$36), whereas fronts and columns are expected to be relatively elongated.\\
It was also an opportunity to isolate the effect of interaction forces on band structure.
For low values of attraction weight (parameter $\gamma$), the repulsion force weight (parameter $\alpha$) seemed to have a tendency to make the nymphs disperse radially as a result of avoidance, leading to poorly aligned cylindric shapes ($\alpha$ to average band's alignment Spearman $\rho=-0.505$, N=36, p $=$ 0.001).
The alignment force weight (parameter $\beta$) would balance this effect by homogeneizing their directions, assembling them in a more aligned crescent-shaped formations ($\beta$ to average band's alignment Spearman $\rho=0.479$, N=36, p $=$ 0.003).
Increasing $\gamma$ strongly overcame the effect of the latter two forces, gathering the individuals in very dense, rather round-shaped formation. ($\gamma$ to average band's area Spearman $\rho=-0.865$, N=36, p $<$ 0.001).
But the conclusion is that without our new local rules, none of the patterns we aimed for emerged.

\subsection{Effect of intermittency and hopping probability on band's shape}

\begin{figure}
    \begin{minipage}[c]{.46\linewidth}
        \centering
        \includegraphics[scale=0.5]{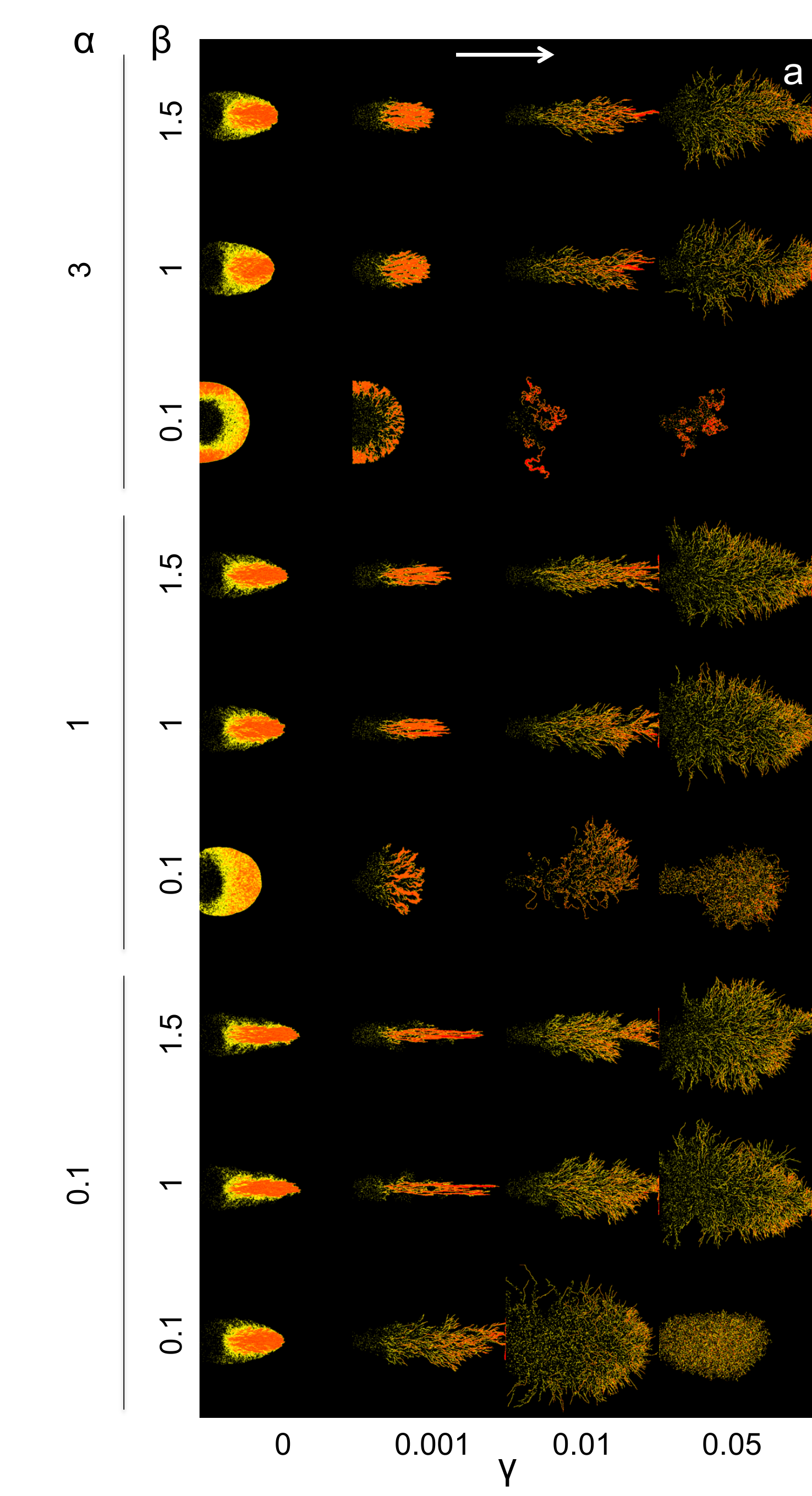}
    \end{minipage}
    \hfill%
    \begin{minipage}[c]{.46\linewidth}
        \centering
        \includegraphics[scale=0.5]{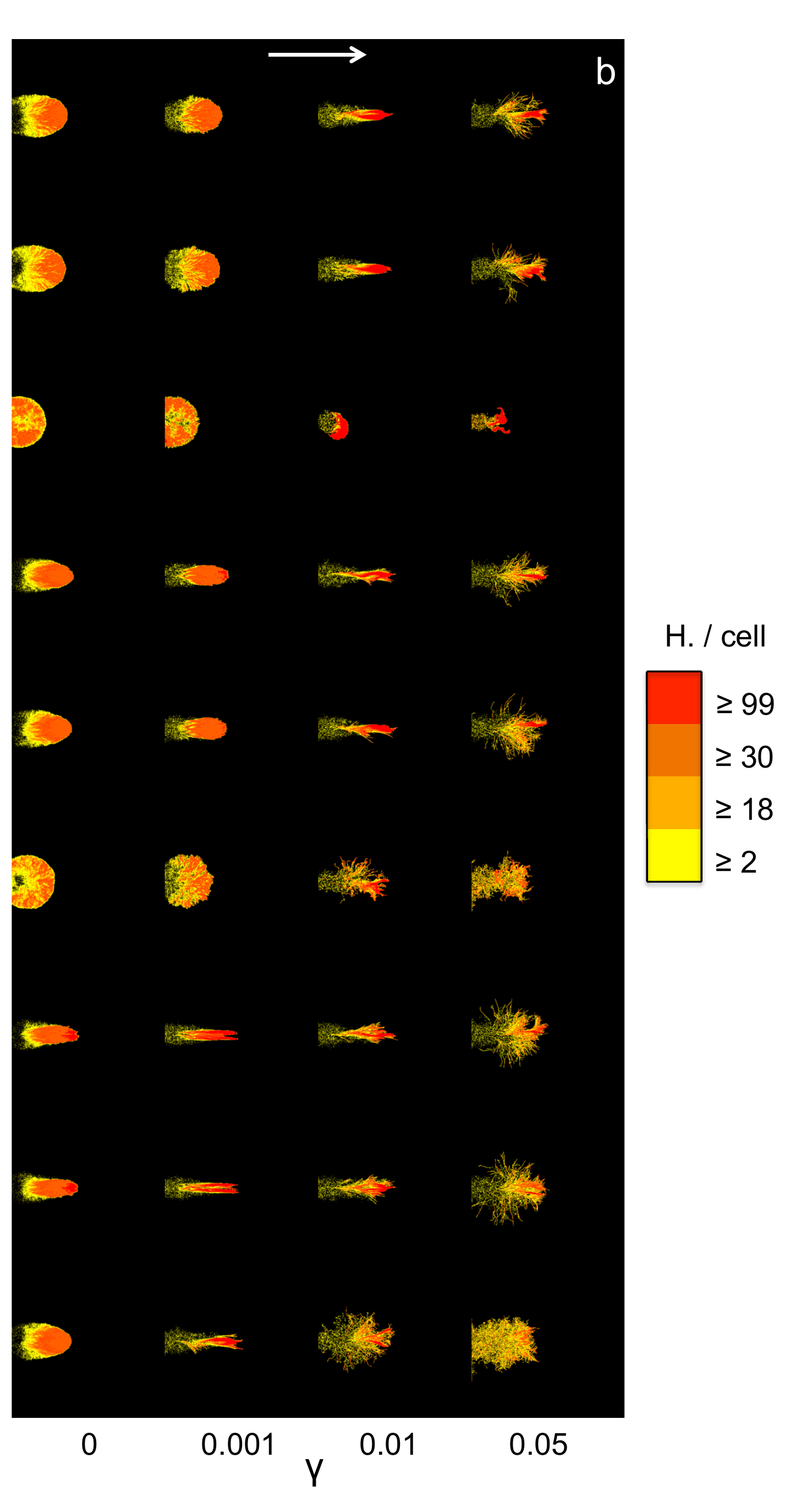}
    \end{minipage}
\caption{Parameter landscapes for the simulated bands after four hours marching when \textbf{(a)} $p_{hc} = 0.2$ and $I=90$ / $p_{res} = 0.5$, in which columnar-shaped bands can be observed, and \textbf{(b)} $p_{hc} = 0.1$ and $I=900$ / $p_{res} = 0.001$, where one can spot frontal shapes.
Left vertical axis is repulsion weight $\alpha$ in the individual interaction force computing, the right vertical one is alignment weight $\beta$ and the horizontal is attraction weight $\gamma$.
Colours describe hopper density (in cell$^{-1}$).
The white arrow indicates the direction of the band at initialization.}
\label{res21}
\end{figure}

After considering the previous results, I systematically explored the parameter values for our new rules for activity intermittency (parameters $I$ / $p_{res}$) and high-density jumping (parameter $p_{hc}$), generating 576 combinations (complete parameter exploration figures available at \url{http://tiny.cc/cy7muy}).
These parameter ranges simulated locust nymphs actively marching for a period of 45 min, after which they would rest for a variable period depending on $p_{res}$, but at least for $I$ seconds.\\
Adding those new rules had a striking effect on the resulting patterns which exhibited a much wider range of shapes, some of which qualitatively looked close to what is expected in actual locust species (examples in figure \ref{res21}).
In a preliminary approach, we focused on identifying the most relevant patterns on a qualitative basis in order to relate them to (i) the frontal formation known to occur in several species, including the APL, and (ii) the columnar formation most notably displayed by \textit{Locustana pardalina}.
We identified frontal shapes as a strip of maximal density near the front of the group, with decaying density in direction of the back of the band, and the columnar shape as a network of dense interwined streams.\\ 

First, strong attraction force ($\gamma=0.01$ and 0.05) mostly resulted in shapes that did not match any known hopper band pattern, regardless of the other parameters values (see figure \ref{res21} for examples).
It appeared that, unlike what was observed in the section \ref{res1lab}, a strong attraction coupled with high crowded hopping probability pushed the band to rapidly disaggregate and spread.
In some parameter ranges, it even resulted in the band being faster than expected, and hitting the eastern boundary of the world.
It was less striking for longer minimal inactivity period ($I$), for which the bands were more coherent, but attraction force gathered them in a spear shape, irrelevant to this study.
However, zero attraction seemed to favour the emergence of frontal shapes for the weakest repulsion, while the 0.001 value displayed several columnar formations for $\alpha = 0.1$ and 1. I focused on these attraction values in the next sections.\\

%

\begin{figure}
    \begin{minipage}[c]{.46\linewidth}
        \centering
        \includegraphics[scale=0.5]{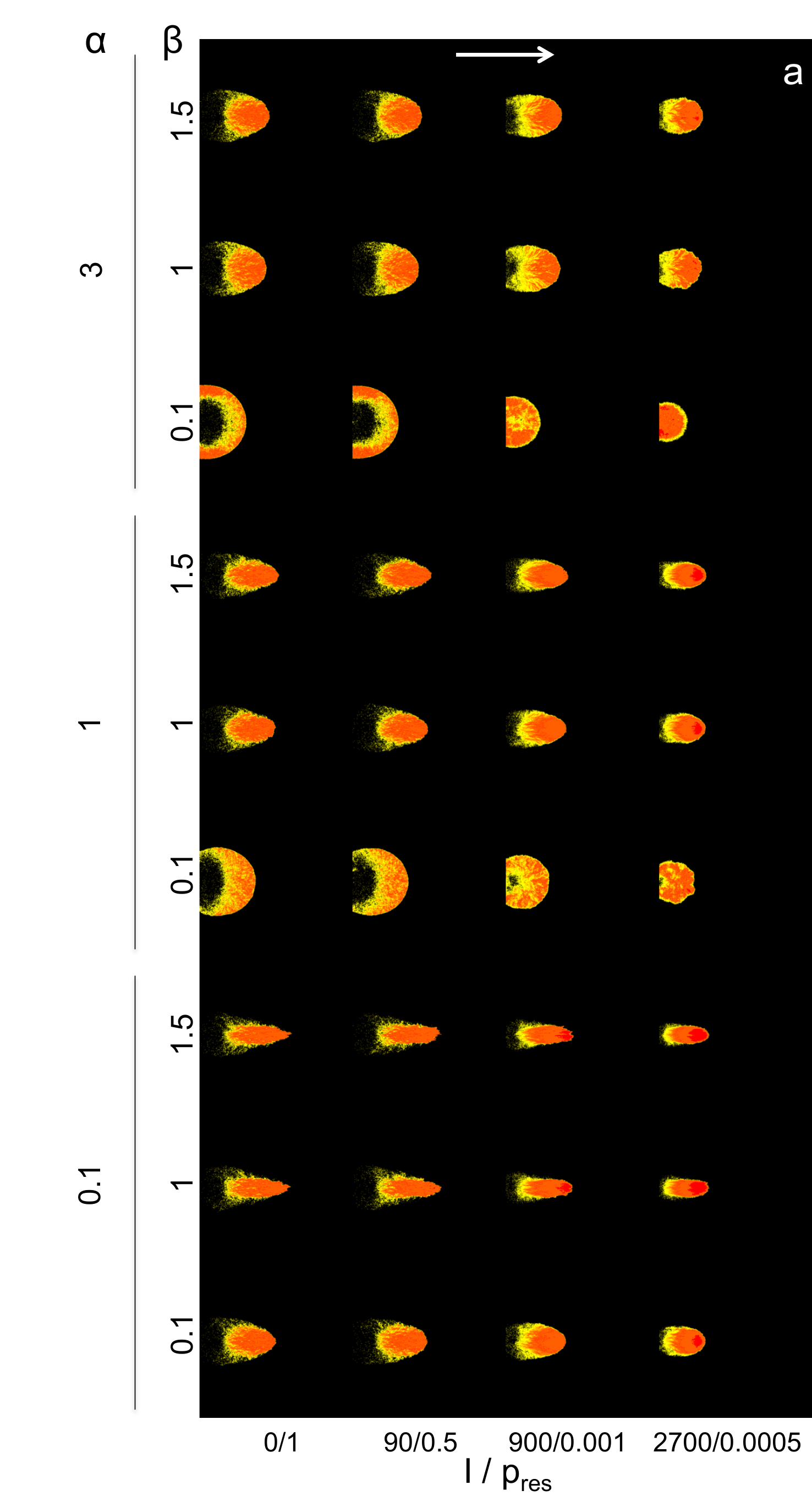}
    \end{minipage}
    \hfill%
    \begin{minipage}[c]{.46\linewidth}
        \centering
        \includegraphics[scale=0.5]{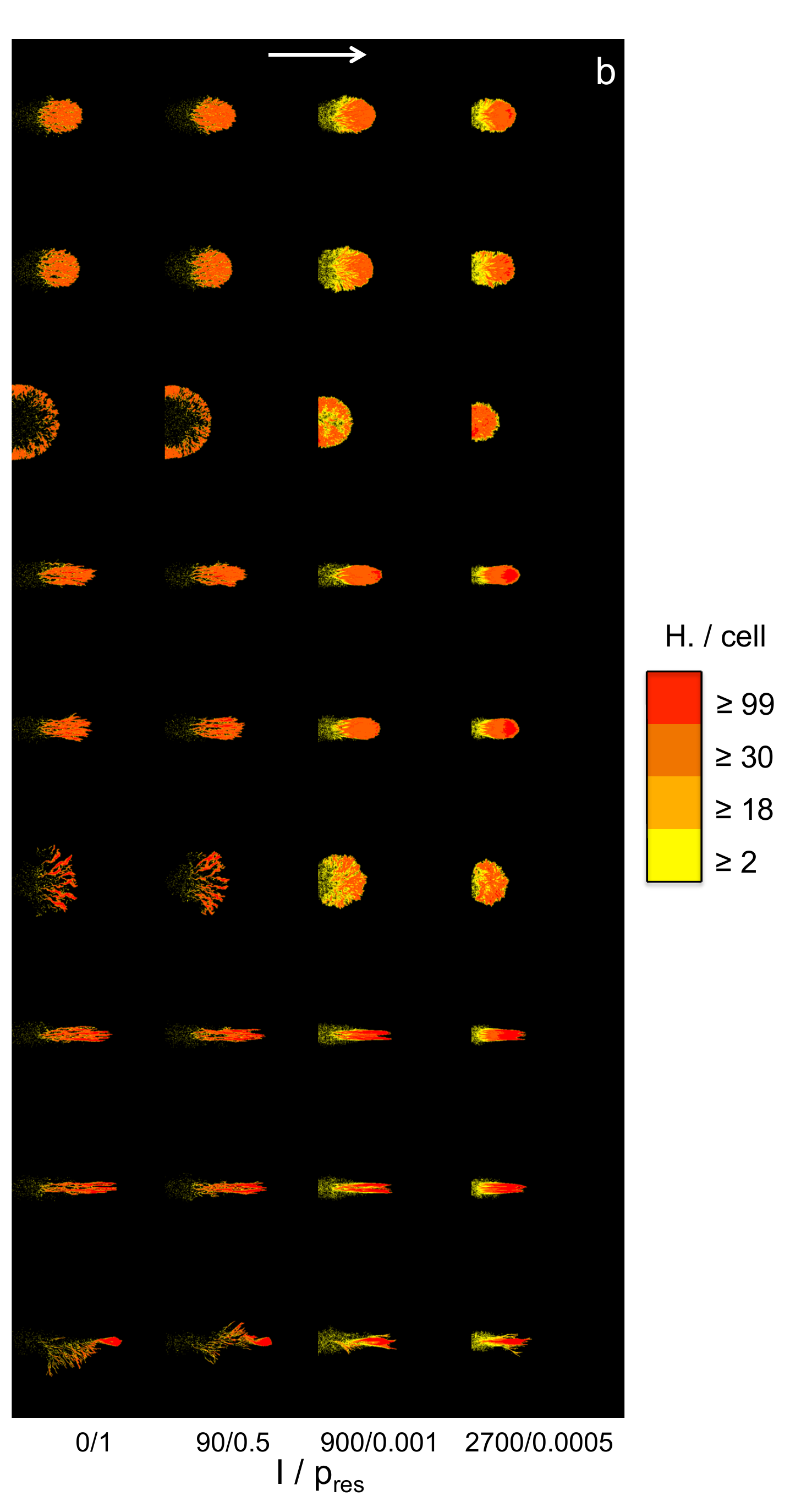}
    \end{minipage}
\caption{Parameter landscapes for the simulated bands after four hours marching for $p_{hc} = 0.1$ and \textbf{(a)} $\gamma = 0$ and \textbf{(b)} $\gamma = 0.001$.
Left vertical axis is repulsion weight $\alpha$ in the individual interaction force computing, the right vertical one is alignment weight $\beta$ and the horizontal are the minimal inactivity period $I$ / probability to resume marching when stopped $p_{res}$ couples.
Colours describe hopper density (cell $^{-1}$).
The white arrow indicates the direction of the band at initialization.}
\label{res22}
\end{figure}

To investigate the role of intermittency on the emergence of each type of pattern, I plotted all the parameter sets for fixed $\gamma$ and $p_{hc}$ values, and qualitatively assessed the effect of the remaining parameters on bands' shape (examples in figure \ref{res22}).
Frontal formations emerged for long $I$; the parameter combination displaying the most of them being $\gamma = 0$ and the $I/p_{res} = 2700/0.0005$ couple (see figure \ref{res22}a).
For $\gamma=0.001$, frontal shapes appeared only for the intermediate value for $\alpha$ and the longest $I$.\\
Columnar formations emerged only for $\gamma=0.001$ and appreared more clearly for short to zero $I$. 
It also looked like the lower the $\alpha$, the closer to what can be observed in brown locust bands (see figure \ref{res22}b).\\
Regardless of $\gamma$ values, the strongest $\alpha$ did not allow the leading edge to be dense enough for a frontal shape nor the streams to be distinguishable enough for a columnar formation.
Moreover, the weakest alignment also resulted in the bands spreading radially in this parameter range.
In conlusion, a pause period of 15 to 45 minutes was necessary for the emergence of fronts, but not of columns, which emerged for short, to absent resting times.\\

%

\begin{figure}
    \begin{minipage}[c]{.46\linewidth}
        \centering
        \includegraphics[scale=0.5]{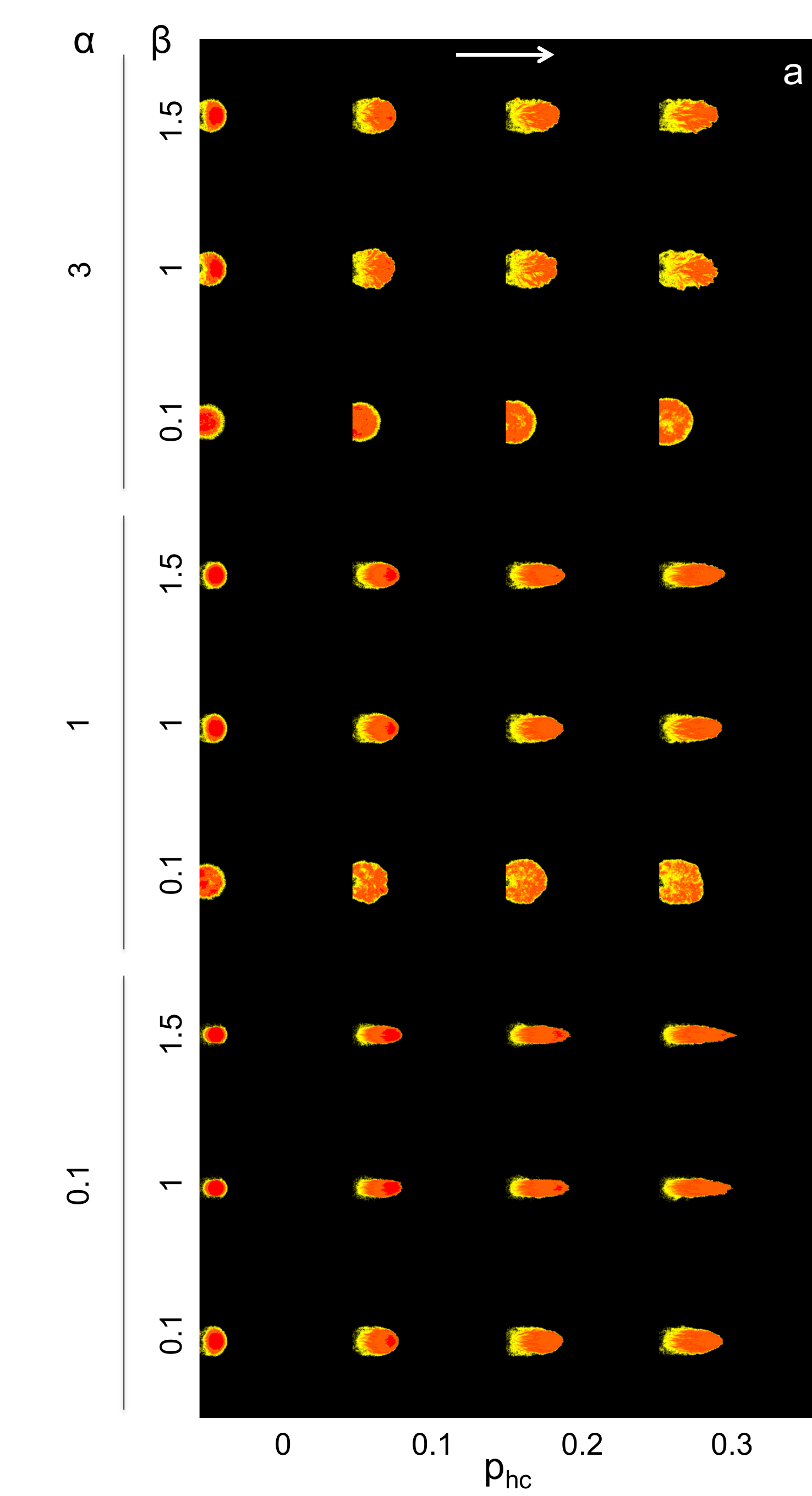}
    \end{minipage}
    \hfill%
    \begin{minipage}[c]{.46\linewidth}
        \centering
        \includegraphics[scale=0.5]{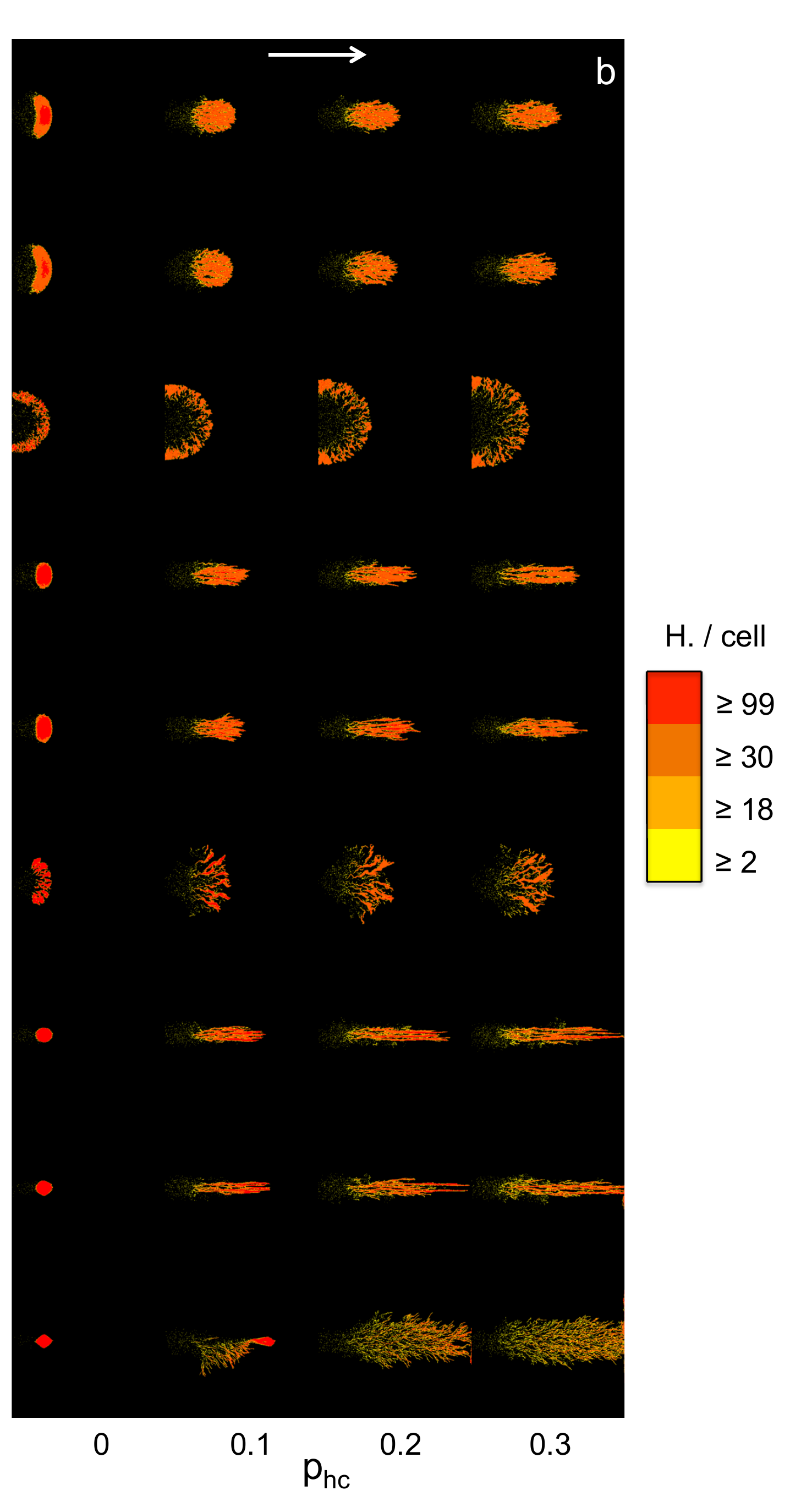}
    \end{minipage}
    \begin{minipage}[c]{.46\linewidth}
        \centering
        \includegraphics[scale=0.5]{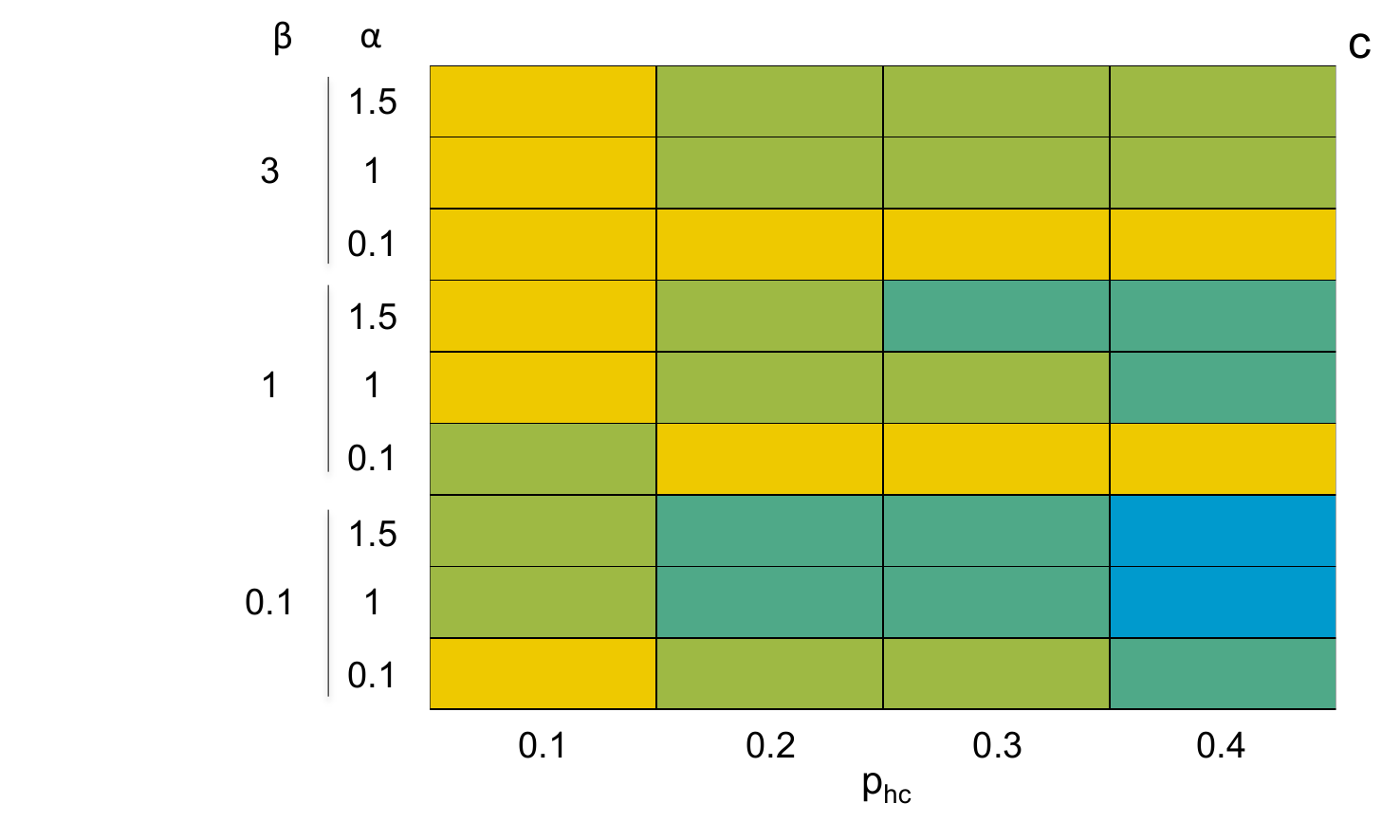}
    \end{minipage}
    \hfill%
    \begin{minipage}[c]{.46\linewidth}
        \centering
        \includegraphics[scale=0.5]{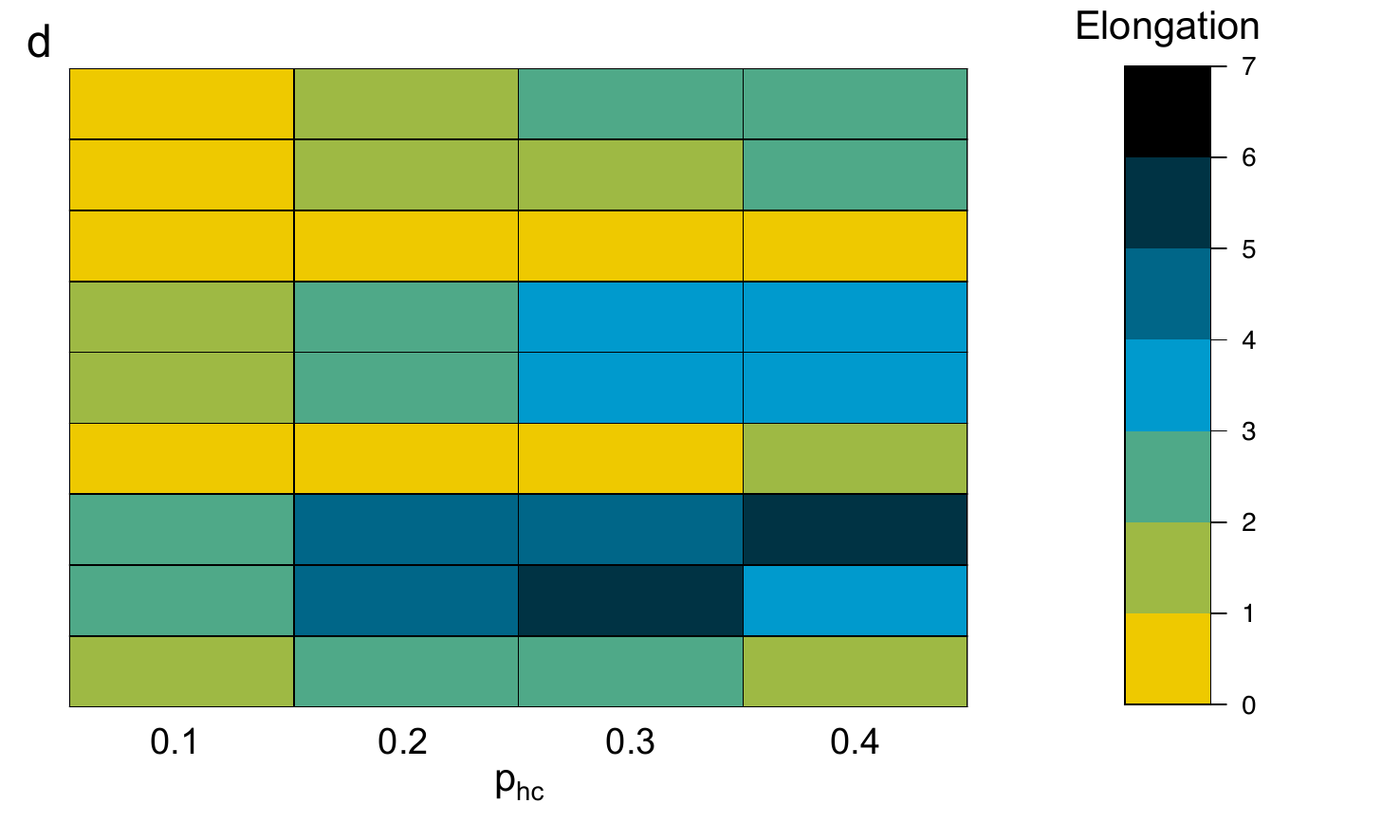}
    \end{minipage}
\caption{Parameter landscapes for the simulated bands after four hours marching, and associted elongation, with \textbf{(a, c)} $\gamma = 0$ and $I/p_{res} = 2700/0.0005$, and \textbf{(b, d)} $\gamma = 0.001$ and $I/p_{res} = 0/1$.
Left vertical axes are repulsion weight $\alpha$ in the individual interaction force computing, the right vertical ones are alignment weight $\beta$ and the horizontal is crowded hopping probability $p_{hc}$.
Colours describe hopper density (cell$^{-1}$).
The white arrow indicates the direction of the band at initialization.
Note that elongation can be misleading in the case of cylindric shapes.}
\label{res23}
\end{figure}

Concerning the role of the crowded hopping probability, I plotted all the parameter sets for fixed $\gamma$ values and $I/p_{res}$ couples and qualitatively assessed the effect of the remaining parameters on band shape (examples in figure \ref{res23}).
When present, increasing $p_{hc}$ seemed to stretch the shape of the bands without strickingly changing the type of formation, and it was quite clear that the higher the probability to jump when crowded, the higher the elongation of the band. Indeed, $p_{hc}$ was positively correlated to elongation (Spearman $\rho=0.209$, N=576, p $<$ 0.001) and especially to band's length (Spearman $\rho=0.729$, N=576, p $<$ 0.001). 
This effect seemed to be tempered by increasing repulsion, which was coherent with $\alpha$ being positively correlated to elongation (Spearman $\rho=0.274$, N=576, p $<$ 0.001).\\
For the longer $I$ and no attraction force ($\gamma = 0$ and $I/p_{res} = 2700/0.0005$), frontal formations have emerged for $p_{hc} = 0.1$ and 0.2 only. The elongating effect of $p_{hc}$ seemed to be needed for their presence, but only until a certain value beyond which the dense front feature was not displayed anymore.\\
For $\gamma = 0.001$ and $I/p_{res} = 0/1$, columnar formations have emerged at all non-zero $p_{hc}$, but for its highest values the bands were faster than expected and had hit the eastern boundary at eight hours marching.
Yet, locust having different probabilities to hop according to density was a necessary condition for the emergence of both patterns.\\

\begin{figure}
	\begin{minipage}[c]{.46\linewidth}
        \centering
        \includegraphics[scale=0.4]{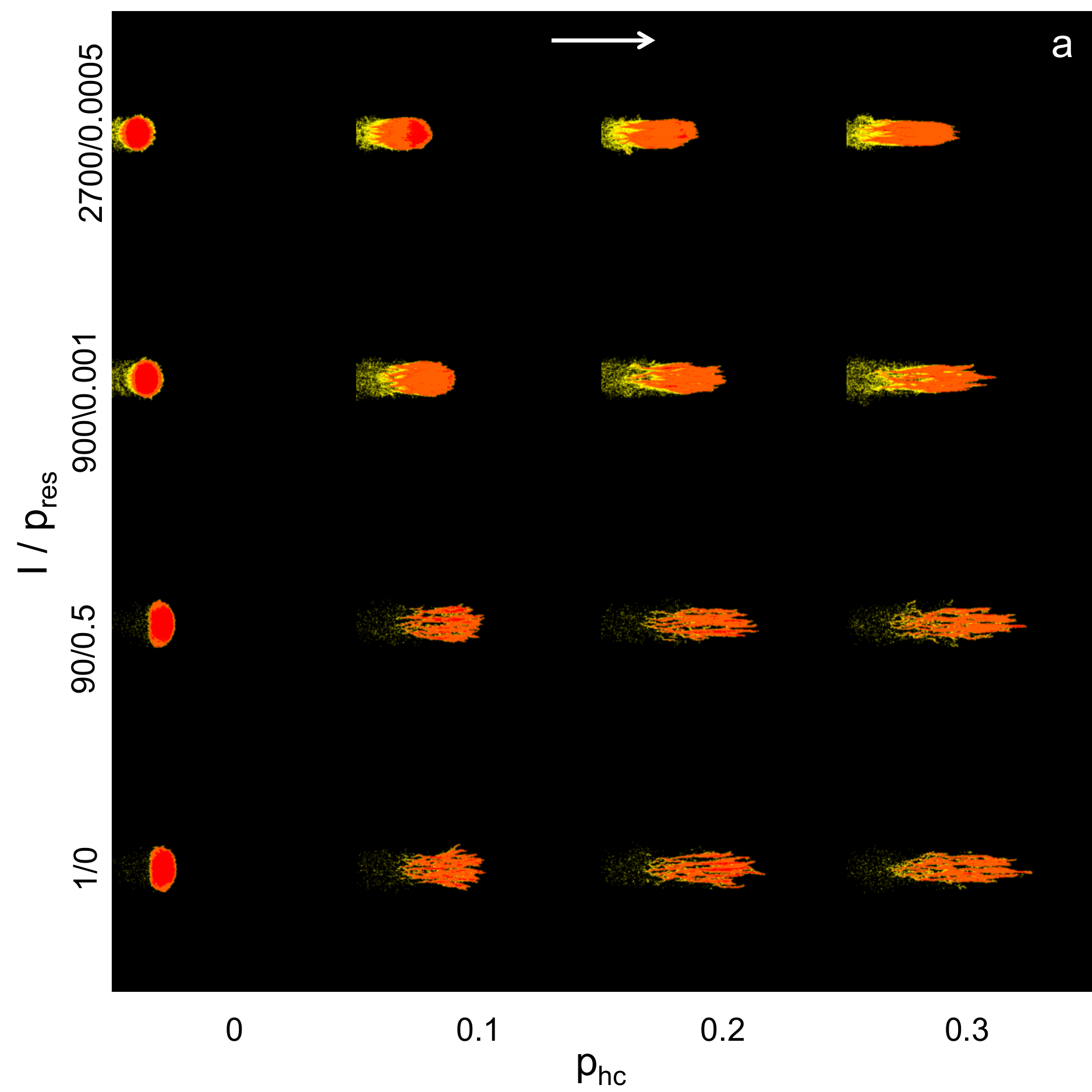}
    \end{minipage}
    \hfill%
    \begin{minipage}[c]{.46\linewidth}
        \centering
        \includegraphics[scale=0.4]{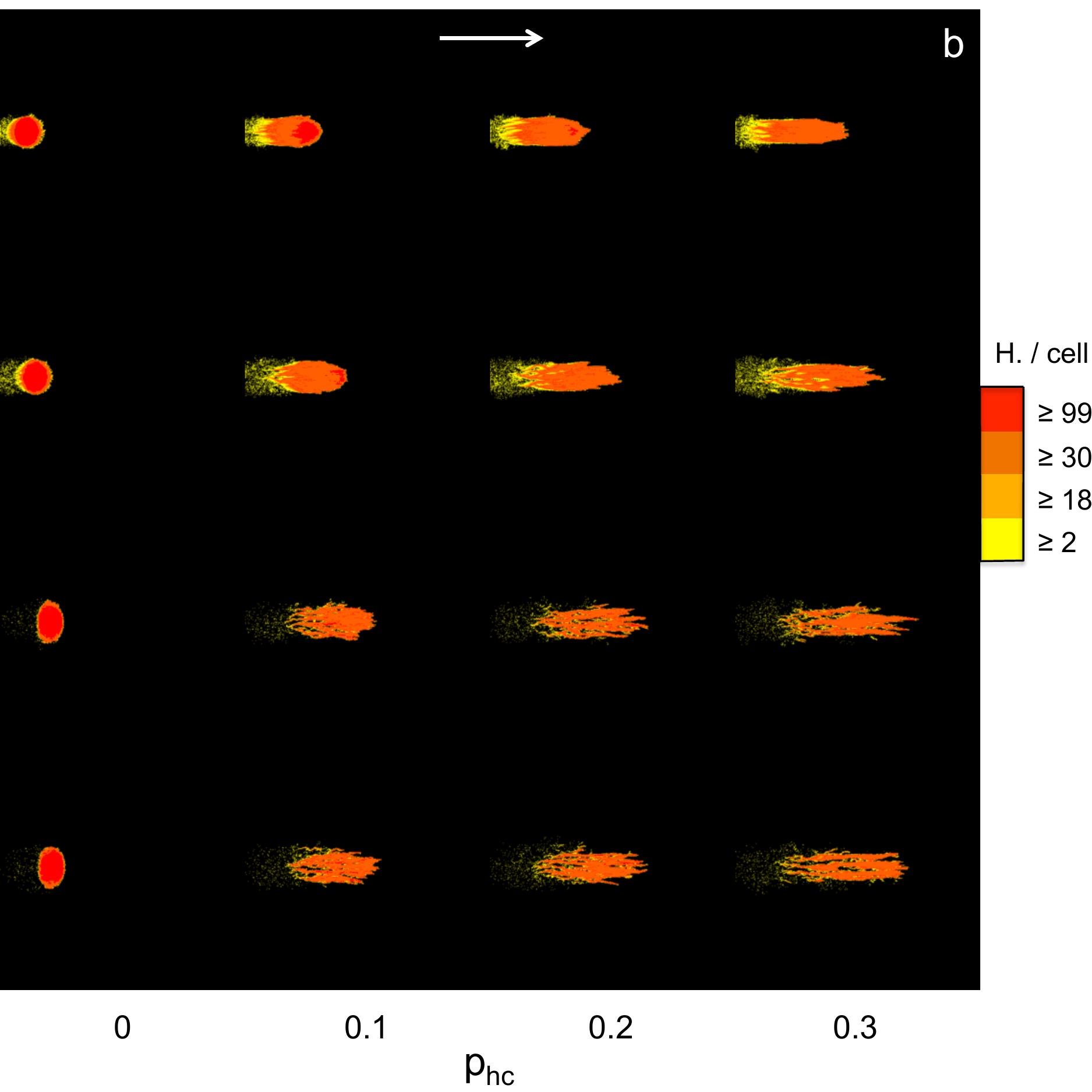}
 	\end{minipage}
\caption{Parameter landscape for the simulated bands after four hours marching when (\textsf{a}) $\alpha = 1$, $\beta = 1$ and $\gamma = 0.001$, and (\textsf{b}) $\alpha = 1$, $\beta = 1.5$ and $\gamma = 0.001$.
Vertical axis is minimal activity period $I$ and $p_{res}$ probability to resume marching couples, and the horizontal one is high-density hopping probability $p_{hc}$.
Colours describe hopper density (cell$^{-1}$).
The white arrow indicates the direction of the band at initialization.
The shapes were not qualitatively different after eight hours marching.}
\label{res24}
\end{figure}

Interestingly, there were two interaction forces combinations that could result in the emergence of both the patterns we were interested in, according to $p_{hc}$ and $I$ / $p_{res}$ values (displayed in figure \ref{res24}).
Consistently with previous results, for these interaction weight combinations and for a given $I$ / $p_{res}$ couple, a lower $p_{hc}$ leaded the bands towards a frontal shape, while a higher $p_{hc}$ stretched the band towards a columnar shape. On the contrary, for a fixed $p_{hc}$, a shorter $I$ resulted in the emergence of columnar patterns and the longer ones of frontal patterns.
Therefore, hopper bands could share exactly the same combination of interaction weights and yet display two completely different patterns of collective motion according to the potential differences in their activity rhythms and/or the threshold of neighbouring conspecifics to trigger an escape response. 

\subsection{Comparison with empirical data on the Australian plague locust}

\begin{figure}
\centering
\includegraphics[scale=0.6]{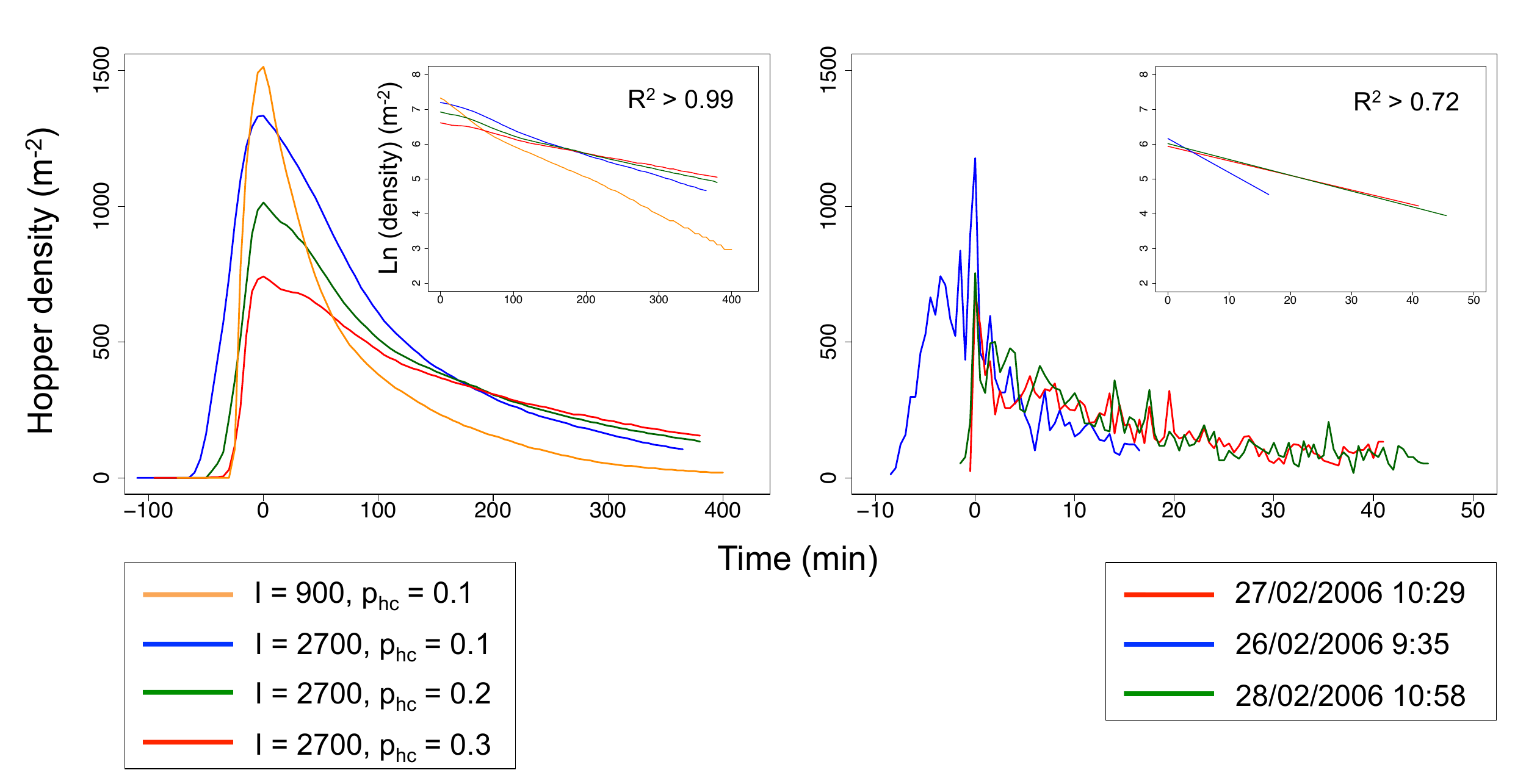}
\caption{Panel \textsf{(a)} shows the density profiles averaged over fifty simulations of parameters $\alpha = 0.1$, $\beta = 1.5$ and $\gamma = 0$, and different values for $I/p_{res}$ and $p_{hc}$.
95\% confidence intervals were too small to appear clearly on the plot ($<$2.9 m$^{-2}$).
In the inset is displayed the average densities after the peak occurence in a semi-logarithmical scale and the minimum R$^2$ from their linear regressions. 
Panel \textsf{(b)} is the density profiles measured from three bands of APL in the wild.
The inset shows the logarithmical regressions of densities after peak occurence and their minimal R$^2$.
The x-axes origins have been made relative to the occurrence of the density peak.
The field bands were $\approx$ 50 to 200 meters long, and simulated bands had an average length of 161.7 m $\pm$ 20.2 (n$=$24) after four hours marching.
The density profiles from differents combinations of interaction weights of the same $I$ and $p_{res}$ values were hardly distinguishable.}
\label{res3den}
\end{figure}

APL hopper bands longitudinal density profiles from field data \citep{buhl2012using} exhibited a sharp peak at the most frontal position (average time of occurrence of peak density after the first locust observed : 4.58 min $\pm$ 3.26 SD; n$=$6) and an exponential decrease from the peak to the rear of the band (logarithm regression slope $< -0.0332$; R$^2 > 0.72$; n$=$6).
To assess whether simulations behaved the same way, I qualitatively selected the parameter sets that resulted in a frontal-shape band (24 candidates) and looked for the same features in their density profiles, averaged on the fifty replicates.\\
Profiles from these simulations displayed the same marked peak at the front and decay in direction of the rear of the band (see figure \ref{res3den}).
Indeed, the average hopper density at peak occurence was 1132.41 m$^{-2}$ $\pm$ 269.36 SD, whereas it was 410.88 m$^{-2}$ $\pm$ 79.26 SD two hours afterwards.
The inset in figure \ref{res3den}a also shows that the decay was exponential (logarithm regression $R^2 >$ 0.987; n$=$24), which matched the field measures.\\
Yet, compared to field hopper bands, the increase in density preceding the peak was less steep in simulations (average time of occurrence : 44.79 min $\pm$ 13.47; n$=$24) and after-peak density was decaying slower (logarithm regression slope $> -0.0107$; n$=$24).
It could be related to differences in band size (difficult to measure in empirical studies, but estimated to be smaller than half a million locusts) and/or average speed (not measured in these field studies and known to strongly vary with temperature and weather conditions).\\
Concerning the effect of intermittency and crowded hopping on the density profiles, it appeared that for the same $p_{hc}$ value, a shorter minimal inactivity period resulted in the peak being more intense, and occuring earlier after the first hopper observed.
Moreover, for a fixed minimal inactivity period $I$, increasing $p_{hc}$ seemed to reduce the intensity of the peak, but also shorten its time of occurence (figure \ref{res3den}a).
Therefore, the most field-consistent simulated frontal formations were obtained for 45 minutes activity and 15 minutes minimal inactivity periods, which was what \cite{simpson1981oscillation, SIMPSON1986480} previously depicted.\\
Furthermore, the exponential decay after the peak could be an expected feature with the probability to resume marching being constant over time. As the amount of time spend being inactive got larger, the number of individuals likely to resume marching would be increasing, possibly leading to the survival curve-like decay in hopper's density observed at the rear of the band in both experiment and simulations.\\

\begin{figure}
\centering
\includegraphics[scale=0.6]{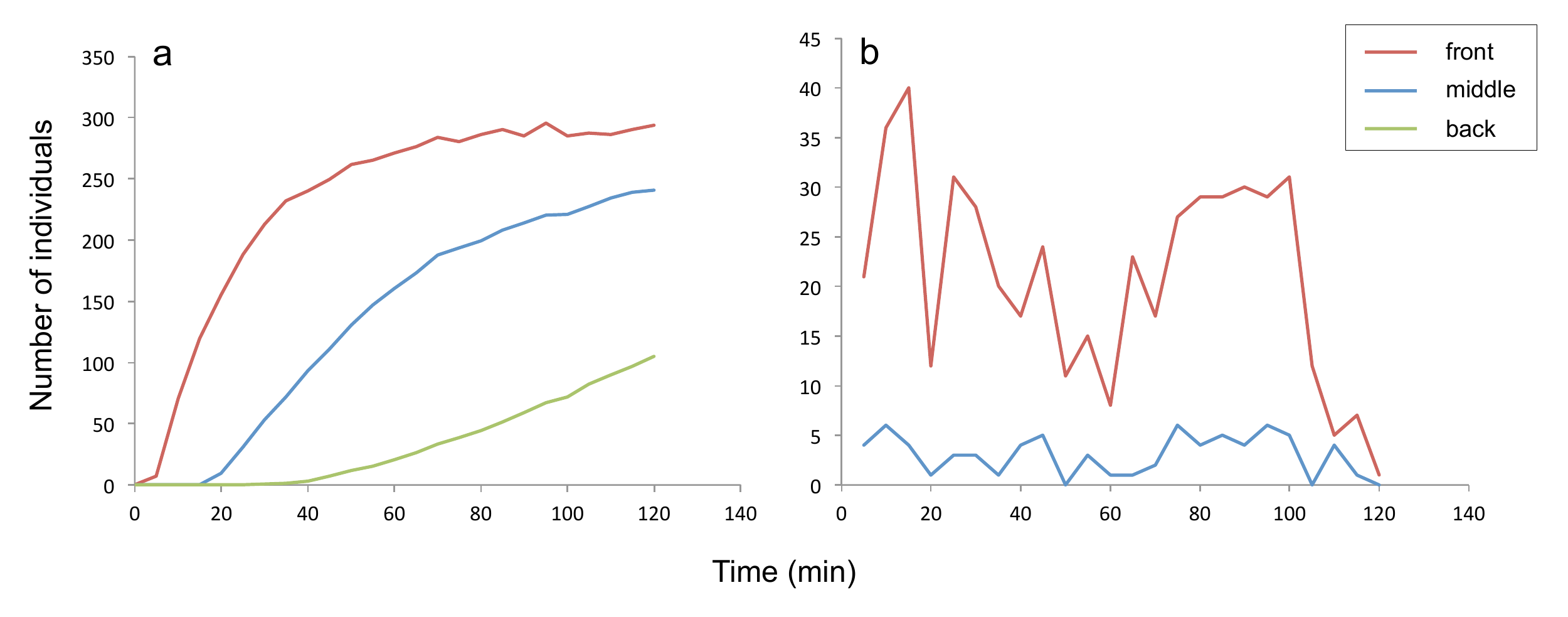}\\
\caption{Panel \textsf{(a)} shows the number of individuals of each tag crossing the measure line according to time, averaged over fitfy simulations of parameters $\alpha = 0.1$, $\beta = 1.5$, $\gamma = 0$, $p_{hc} = 0.1$ and $I/p_{res} = 900/0.001$.
The origin of the time axis corresponds to fifty minutes marching. 
95\% confidence intervals were too small to appear clearly on the plot ($<$3.3 individuals). 
Panel \textsf{(b)} is the number of APL nymphs in a one square-meter area at the middle of the front of the band according to time (summed over four different observators placed 5m apart).
The origin of the time axis corresponds to 3h20min marching.
}
\label{res3tag}
\end{figure}

In the paint marking experiment on simulations, individuals initially from the front and the middle of the band started to cross the measure line together only ten minutes after the first locust counted (figure \ref{res3tag}).
Locust nymphs initially from the back were observed approximately thirty minutes later.
In the empirical data, individuals initially from the front and from the back appeared together from the very first five minutes counting. 
Note that on the field, the band have been marching for several hours before the experiment started, against only fifty minutes in simulations.\\
In both experiments and simulations, individuals initially from different positions in the band occured together at the front after a few time marching, matching our prediction for the global-level behavior of a band displaying marching activity intermittency, and the rolling structure associated.

\newpage
\section{Discussion}

In this study, we addressed the role of speed heterogeneity in the emergence of frontal and columnar formations in locust hopper bands, following the framework of Self-Organization in biological systems.
Based on literature and observations, we hypothetized that marching activity intermittency and density-dependent hopping
could play a role in this emergence. 
We then formulated these hypotheses in a three-zone SPP model variant in which individuals alternated between marching and resting, and were more likely to hop when crowded.
To predict the global patterns that could emerge from these rules, we simulated half a million locust nymphs marching during eight hours, in fifty replicates of more than five hundred parameter combinations; a data set size that is hardly matched in the study of collective motion. 
This challenge was overcome by implementing the model in the CUDA C++ language using parallel computing, and running the simulations on a large amount of GPGPUs in the University of Adelaide's high performance computer Phoenix.\\

Our model successfully predicted the emergence of both columnar and frontal formations.
Moreover, the simulated frontal shapes exhibited the same density structure as actual \textit{C. terminifera} hopper bands \citep{buhl2011group}. Interestingly, the best fits were obtained for an alternance between 45 min activity and 15 min minimal pause; as previously suggested by \cite{simpson1981oscillation} and \cite{SIMPSON1986480} on the basis of behavioural experiments on \textit{L. migratoria}.\\
The crowded hopping behavior was a necessary condition to the emergence of both our patterns of interest. When present, it appeared to have an elongating effect on the bands, which could defavouring the formation of a dense front when probability was too high, but without affecting the emergence of columns.\\
Both experiments and simulations validated our global predictions for the existence of a rolling structure linked to marching activity intermittency in hopper bands exhibiting frontal formations, for the emergence of which it was a necessary condition.
Inversely, columns emerged for short to zero minimal resting periods only.
This suggests that locust nymphs should spend more time in an active state in columnar configations than in frontal ones.
This new prediction would be interesting to test in a comparative study of the activity rythms of \textit{L. pardalina} and \textit{C. terminifera}.
The usage of miniature harmonic radar transponders and millimetre-wave radars should allow such measurements in future field experiments.\\

Two particular interaction forces weight combinations ($\alpha = 0.1$, $\beta = \{1;1.5\}$, $\gamma = 0.001$) could result in the emergence of both patterns, according to the minimal inactivity period and the high-density hopping probability values. 
Specifically, for $p_{hc}=0.1$, increasing the period of inactivity could lead a columnar formation to change into a frontal one, and conversely.
Beside being a good example of what a spectacular source of variability Self-Organization can be,
this could explain why previous field observations reported that, in a same species, sometimes the same hopper band, could display several different shapes, ranging from frontal to columnar patterns.
Supposing that such a switch was not linked to external constraints, it could be explained by the intermittency being a dynamic behavior.
Nutritional state influencing collective behaviour have already been observed in non-specific simulations of social animal foraging in spatially complex environnements \citep{liho17}.
For example, locusts could shorten or enlarge their minimal pause period according to their nutritional state and food availability, and if social interaction forces weights of these nymphs were in such a critical parameter range, this change could stretch the band to a columnar formation and vice versa.
One could suspect that, in the case of scarce resources, migrating in frontal formation could possibly be more advantageous, as it would reduce the chance to miss food patches by covering a wider area than columns.\\

This study also showed the need for a statistical sensitivity analysis adapted to non-deterministic models.
Such a tool would ease the understanding and interpretation of parameters effects, especially their interactions, on the shape of the patterns.
But, although the proxys for band shape I chose to compute were useful as a qualitative exploration, they were not discriminative enough of the formations we were interested in, which would be a necessary feature for any sensitivity analysis.
The perimeter to area ratio could be a good candidate, as it should be significantly larger in columnar than in frontal shapes.
Also, gathering this kind of data from the field could be an interesting challenge as it will require measurements to be taken from a birds eye point of view.
In future field experiments, the usage of drones and photogrammetry should allow more accurate measures on bands shape, especially the detection and quantification of the densest parts.\\

During this investigation, I gathered other quantitative measures remaining to be analysed, notably a full trajectometry of the data set.
Improved locust hopper bands control relies on spraying a biopesticide (based on the fungus \textit{Metarhizium anisoplae}) in dense swats over vegetation \citep{hunter2001aerial}, requiring the nymphs to feed on it. 
Being able to better understand and predict the trajectory of the band, especially its probability to turn, is a key to optimizing control methods.\\

Finally, another way to improve our understanding and modelling of collective movement could involve studying inter-individual interactions at the neurobiological level.
It has long been suggested that locusts might align with their near neighbours by minimising optic flows in their visual field.
In the future, virtual reality experiments coupled with a neuro-ethological approach might allow us to rethink how interaction rules are truely implemented in animals.

\newpage
\nocite{*}
\bibliography{memoir_M2_arXiv_sub.bib}

\newpage
\appendix

\section{Appendices}
\subsection{Details on model implementation}

Each particle was described by six instantaneous variables, that were $x$ and $y$ spatial coordinates, $\theta_x$ and $\theta_y$ direction vector coordinates, the number of neighbours in its alignment and repulsion range, and its value on the time tracker. At each time-step, they were updated in the following order :
\begin{enumerate}[label=Step \arabic*:]
	\item Updating direction according to equation \ref{eq2}.
	\item Testing for neighbour in $R_r$ and set the hopping probability to $p_h$ or $p_{hc}$ accordingly.
	\item Tossing with the hopping probability to apply or not the $H$ hopping speed bonus to $v_0$.
	\item Testing for activity state. If the time tracker value was between 0 and $-I$, setting speed to zero. If it was below $-I$, tossing with $p_{res}$ to reset the time tracker to $A$.
	\item Computing next position according to equation \ref{eq1}.
	\item Checking if particle still in the world boundaries and if not using boundary as a reflective wall to update position.
	\item Converting the position in grid coordinates. Grid sorting to spot neighbours within $R_a$.
	\item Computing independently the normalized sum of repulsion, alignment and attraction forces according to \ref{eq4}. Plus updating neighbour's number within $R_p$.
	\item Testing for occlusion and summing the different forces accordingly.
	\item Computing next direction according to equation \ref{eq3}.
\end{enumerate}

\subsection{Details on data extraction from simulations}
In the following section, uppercase letters refer to cell-related values, lowercase letters refer to cartesian values, exept for $N$ that just refers to the total number of individuals.\\
I wrote a Perl script to gather all the information needed for these measures at once to restrain the number of output readings to one.\\
Browsing the output file, for each time step, it collected the grid coordinates and converted them in cartesian coordinates thank to their hash code. These values as well as the number of occupied cells $C$ and the number of particles in each one of them $n_{I}$ was temporary stored for computation.
Before changing time step, it computed the coordiates of the cell containing the center of mass (CM) following equation \ref{cm}, it collected the average direction of the band $\bar{\theta}$ and compute its norm $|\bar{\theta}|$ as a proxy of alignment within the band, which is the classical proxy in collective movement studies.\\
\begin{equation}
\label{cm}
\begin{aligned}
\{X_{CM}\ ;\ Y_{CM}\} &= \Bigg\lbrace \frac{\sum\limits_{I=1}^C n_I.X_I}{N}\ ;\ \frac{\sum\limits_{I=1}^C ni.Y_I}{N} \Bigg\rbrace
\end{aligned}
\end{equation}

If the current frame was superior or equal to the time I chose to set the line, a change of coordinates system was performed following eq. \ref{base}, such that the grid was represented in a new base ($\vec{e_x'}$;$\vec{e_y'}$) which unit vector $\vec{e_x'}$ was colinear to the mean direction of the band $\bar{\theta}$. That way, I could choose a fixed $X_{line}'$ on $x'$ axis where to set the transversal area perpendicular to the average direction of the band.
\begin{equation}
\label{base}
\begin{aligned}
X_I' &= \sqrt{X_I^2+Y_I^2} . \cos(\arctan(Y_I,X_I) - \bar{\theta})\\
Y_I' &= \sqrt{X_I^2+Y_I^2} . \sin(\arctan(Y_I,X_I) - \bar{\theta})
\end{aligned}
\end{equation}
From this time until the end of the simulation, at each time step, the sum of the individuals in the cells that had $X_{line}'$ as coordinate was stored.\\

If the current frame was one the two times on which we chose to print out the outputs for model exploration, the maximum and minimum of $x'$ and $y'$ were computed to have an idea of the size of the band. For a more informative measure, the minimum and maximum of $X$ and $Y$ among the cells containing more than five individuals was browsed. I also extracted the alignment vector at this time. These measures were written in an other text file.\\

Concerning tagged individuals, I wrote a separate Perl script based on the same method exept that the counting of hoppers in the $X_{line}'$ cells depended on their tag. If a particle's $x'$ coordinate fell in the counting area, the right tracker was being added one according to individual's tag number. Measures were written in a a separate text file.


Finally, I wrote a perl script for each of the three text file type that browsed them and computed each measure's average, standard deviation and 95\% confidence interval by Markov chain method.
There was one text file per parameter set for the density profiles containing the average density in the measure line accross all replicates, plus standard deviation and confidence interval for each time step of simulation output. 
Concerning measures on band's global shape, a single text file contained each parameter value plus average, SD and CI95 on all replicates for bands' width, length, cohesion and surface after four and eight hours of marching (same time steps as when the band's was printed out).


%
%


\end{document}

%% file: memoir_titlepage.tex
\begin{titlepage} 
	\newcommand{\HRule}{\rule{\linewidth}{0.5mm}} 
	
	\center 
	
	
	\textsc{\LARGE Université Toulouse III - Paul Sabatier}\\[1.5cm] 
	
	\textsc{\Large Ecological Systems Modelling master's internship}\\[0.5cm] 
	
	\textsc{\large School of Agriculture, Food and Wine}\\[0.5cm] 
	
	\textsc{\large The University of Adelaide}\\[0.5cm]
	
	
	\vfill\vfill\vfill
	\HRule\\[0.6cm]
	
	{\huge\bfseries Exploring locust hopper bands emergent patterns using\\[0.2cm]
					parallel computing.}\\[0.5cm]
	\HRule\\[1.5cm]
	
	
	\begin{minipage}{0.4\textwidth}
		\begin{flushleft}
			\large
			\textit{Author}\\
			Adrian \textsc{Bach} 
		\end{flushleft}
	\end{minipage}
	~
	\begin{minipage}{0.4\textwidth}
		\begin{flushright}
			\large
			\textit{Supervisor}\\
			Dr. Jerôme \textsc{Buhl} 
		\end{flushright}
	\end{minipage}
	
	
	
	\vfill\vfill\vfill 
	
	{Handed in on 14 June 2018} 
	
	
	\vfill\vfill
    \begin{minipage}[c]{.46\linewidth}
        \centering
        \includegraphics[scale=0.9]{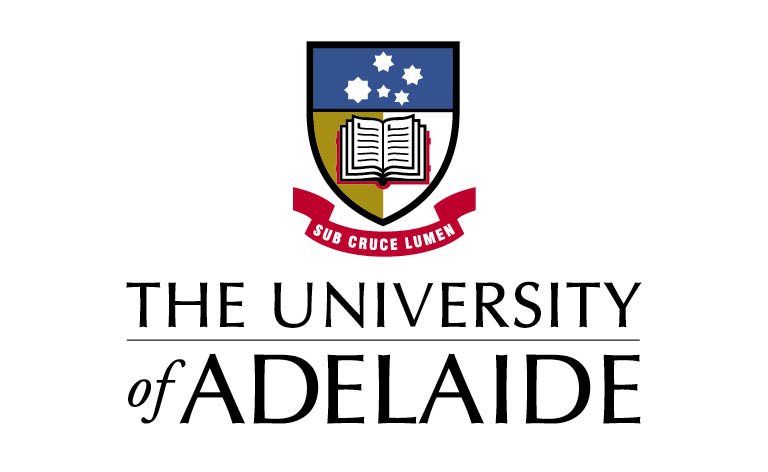}
    \end{minipage}
    \hfill%
    \begin{minipage}[c]{.52\linewidth}
        \centering
        \includegraphics[scale=0.5]{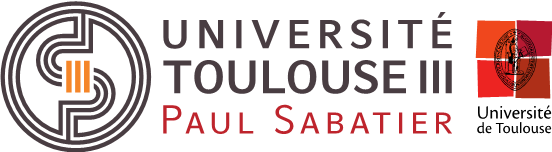}
    \end{minipage}
	 
	
	\vfill
	\noindent
	\tiny This report is an educational exercise that engages in no case the responsibility of the host laboratory.
	
	\vfill 
	
\end{titlepage}

